\documentclass[twocolumn,tighten,times,twocolappendix,numberedappendix]{aastex631}

\usepackage{amsmath}
\usepackage{float}
\usepackage{color}
\usepackage{url}
\usepackage{etoolbox}
\usepackage{graphicx}

\usepackage[flushleft]{threeparttable}

\usepackage{placeins}

\hypersetup{breaklinks=true,colorlinks=true,linkcolor=blue,citecolor=blue,filecolor=blue,urlcolor=blue}
\makeatletter
\patchcmd{\NAT@citex}
  {\@citea\NAT@hyper@{%
     \NAT@nmfmt{\NAT@nm}%
     \hyper@natlinkbreak{\NAT@aysep\NAT@spacechar}{\@citeb\@extra@b@citeb}%
     \NAT@date}}
  {\@citea\NAT@nmfmt{\NAT@nm}%
   \NAT@aysep\NAT@spacechar\NAT@hyper@{\NAT@date}}{}{}

\patchcmd{\NAT@citex}
  {\@citea\NAT@hyper@{%
     \NAT@nmfmt{\NAT@nm}%
     \hyper@natlinkbreak{\NAT@spacechar\NAT@@open\if*#1*\else#1\NAT@spacechar\fi}%
       {\@citeb\@extra@b@citeb}%
     \NAT@date}}
  {\@citea\NAT@nmfmt{\NAT@nm}%
   \NAT@spacechar\NAT@@open\if*#1*\else#1\NAT@spacechar\fi\NAT@hyper@{\NAT@date}}
  {}{}

\usepackage{array}
\newcolumntype{C}[1]{>{\centering\let\newline\\\arraybackslash\hspace{0pt}}m{#1}}

\usepackage{xspace}

\newcommand{\orcidicon}{\includegraphics[width=0.26cm]{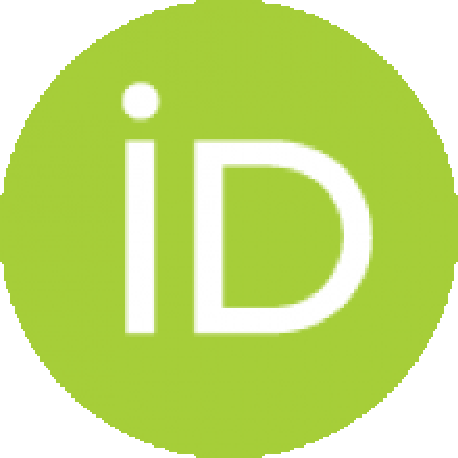}}

\newcommand{\orcidauthor}[1]{\href{https://orcid.org/#1}{\orcidicon}}

\def\araa{ARA\&A}
\def\apj{ApJ}
\def\apjl{ApJL}
\def\apjs{ApJS}
\def\ao{Appl.~Opt.}

\def\aap{A\&A}
\def\aapr{A\&A~Rev.}
\def\aaps{A\&AS}

\def\mnras{MNRAS}

\def\pasj{PASJ}

\def\ssr{Space~Sci.~Rev.}

\def\nat{Nature}

\def\nar{New~Astro.~Rev.}

\newcommand{\ionic}[2]{#1$\,${\scshape{#2}}\xspace}

\def\arcdeg{\hbox{$^\circ$}}

\makeatletter
\patchcmd{\frontmatter@RRAP@format}{(}{}{}{}
\patchcmd{\frontmatter@RRAP@format}{)}{}{}{}
\renewcommand\Dated@name{}
\makeatother

\shorttitle{IGR\,J12580+0134: The Nearest TDE}
\shortauthors{Danehkar}
\graphicspath{{./}{figures/}}

\begin{document}

\title{IGR\,J12580+0134: The Nearest Tidal Disruption Event and its Faint Resurrection}



\correspondingauthor{A.~Danehkar}

\author[0000-0003-4552-5997]{A.~Danehkar}
\affiliation{Eureka Scientific, 2452 Delmer Street, Suite 100, Oakland, CA 94602-3017, USA; \href{mailto:danehkar@eurekasci.com}{danehkar@eurekasci.com}}




\date[]{\textit{\footnotesize Received 2025 January 20; revised 2025 April 16; accepted 2025 April 16; published 2025 June 5}}

\begin{abstract}
Tidal disruption events (TDEs) are X-ray and gamma-ray radiations emerging from the tidal disintegration of a star or substellar object that passes too close to a supermassive black hole (SMBH) at the center of a galaxy. In November 2010, a TDE designated as IGR\,J12580+0134 occurred in the galaxy NGC\,4845, and it was traced by follow-up \textit{XMM-Newton} observations in January 2011. To identify a further TDE based on the radio outburst cycle, we requested \textit{NICER} monitoring observations for nearly a year beginning in March 2023, which we studied here along with the previous \textit{XMM-Newton} observations. We analyzed X-ray brightness changes using hardness analysis and principal component analysis (PCA), and conducted spectral analysis of the source continuum. The \textit{NICER} observations revealed the presence of some X-ray flares during March--June 2023 that were much fainter than the TDE observed using \textit{XMM-Newton} in 2011. The PCA component that mainly contributes to the X-ray outbursts during the TDE is a heavily absorbed power-law continuum emission, whereas there is a small contribution from collisionally ionized plasma in the soft excess, likely from a colliding wind or jet.  Similarly, the PCA of the \textit{NICER} data relates the X-ray flares to a power-law spectrum, albeit with a much lower absorbing column, and partially to soft collisional plasma. The faint X-ray flares captured by \textit{NICER} could be associated with extremely weak accretion onto the SMBH resident in this galaxy and thus potentially a low-luminosity active galactic nucleus.
\end{abstract}


\keywords{
\href{https://astrothesaurus.org/uat/1696}{Tidal disruption (1696)};
\href{https://astrothesaurus.org/uat/17}{Active galaxies (17)};
\href{https://astrothesaurus.org/uat/16}{Active galactic nuclei (16)};
\href{https://astrothesaurus.org/uat/159}{Black hole physics (159)};
\href{https://astrothesaurus.org/uat/1822}{X-ray sources (1822)};
\href{https://astrothesaurus.org/uat/1944}{Principal component analysis (1944)}
\vspace{4pt}
}


\section{Introduction}
\label{ngc3783:introduction}



It is widely accepted that many galaxies host supermassive black holes (SMBHs) at their centers \citep[see reviews by][]{Kormendy1995,Ho2008,Kormendy2013,Graham2016}. The most compelling evidence for SMBHs currently stems from X-ray observations of active galactic nuclei (AGN), which exhibit luminous X-ray flares and iron K$\alpha$ line emission. However, only a small portion of SMBHs are active, and the rest remain quiescent owing to weakly accreting or non-accreting SMBHs as suggested by black hole evolution models \citep[e.g.,][]{Marconi2004,Kormendy2001}. The direct detection of an SMBH in a non-active galaxy is achievable by observing an X-ray brightening burst known as a tidal disruption event \citep[TDE; see reviews by][]{Saxton2020,Alexander2020,Gezari2021}. This phenomenon occurs when a star rips apart as it approaches too close to a massive black hole \citep{Hills1975}. Approximately half of the mass of the destroyed star falls into the black hole event horizon, resulting in bright X-ray and $\gamma$-ray bursts \citep{Rees1988}. A portion of the accreted mass may also be transformed into an outflow \citep{Evans1989}, potentially producing a relativistic jet, called a tidal disruption flare, as seen in late-time radio observations \citep[e.g.,][]{Zauderer2011,Cenko2012,Brown2017,Mattila2018}.

A TDE, later called IGR\,J12580+0134, occurred in November 2010, and it was observed in January 2011 \citep{Walter2011} using the International Gamma-Ray Astrophysics Laboratory satellite \citep[\textit{INTEGRAL};][]{Winkler2003}. It was located in the nucleus of NGC\,4845, which is a nearby spiral galaxy classified as Seyfert 2 \citep{Ho1995,Veron-Cetty2006} with a redshift distance of $18.535$\,Mpc \citep{Lianou2019} or $18.0 \pm 3.5$\,Mpc \citep{Yu2022}. Follow-up observations with \textit{XMM-Newton}, \textit{Swift}, and \textit{MAXI} indicated that the TDE likely involved the disruption of an object with a mass of $14$--$30$ Jupiter by a massive black hole of $\sim 3 \times 10^5 M_{\odot}$ \citep{Nikolajuk2013}. A variable compact radio source was identified in NGC 4845 through the Continuum Halos in Nearby Galaxies-EVLA Survey (CHANG-ES), suggesting the presence of an AGN in this galaxy \citep{Irwin2015}. \textit{XMM-Newton} observations were well fitted by a soft X-ray component with an unabsorbed blackbody temperature of 0.33 keV and a hard X-ray excess using a power-law model with $\Gamma = 2.2$ \citep{Nikolajuk2013}, which could originate from thermally diffuse plasma and the accretion corona close to the SMBH \citep{Lei2016}, respectively. Radio observations, conducted roughly a year after the X-ray peak, revealed a late 1.6-GHz radio peak between December 2011 and March 2012 \citep{Irwin2015,Perlman2022}.  Again, another radio brightening detected in the \textit{S}- and \textit{L}-band was identified in May 2016, although X-ray observations were not collected several months prior to this radio peak, and the source was found to be faint in \textit{Swift}/XRT observations in March and May 2016 and December 2017 \citep{Perlman2022}. 

Based on the 5.5-year cycle of the two radio peaks, a future TDE was expected to occur in 2022. However, no X-ray bursts were found in the follow-up \textit{XMM-Newton} and \textit{NuSTAR} observations collected in June 2021 and May 2022, respectively. In particular, a partial tidal disruption of a star reducing its mass ($M_{\star}$), by a super-Jupiter in the case of IGR\,J12580+0134, may decrease the escape velocity ($v_{\rm esc}=\sqrt{2G M_{\star}/R_{\star}}$) and increase the tidal radius \citep[$r_{\rm t}=R_{\star} (M_{\rm BH}/ M_{\star})^{1/3}$;][]{Rees1988} under simple assumptions. However, this process also depends on other physical characteristics, such as the polytrope ($\gamma$), which makes high-mass stars ($\gamma=4/3$) more easily disrupted than low-mass stars \citep[$\gamma=5/3$;][]{Mainetti2017}. The eccentricity of the stellar orbit can also modify fallback dynamics \citep{Cufari2022a}. Moreover, for late-type giant stars with significant tidal heating, the tidal radius may increase \citep{Sharma2024}, whereas this may not be plausible in compact stars. Hydrodynamic simulations have demonstrated that a partial TDE can modify orbital energy and period \citep{Guillochon2013}. In addition, the orbital period $T\propto \langle \rho \rangle^{-1/2}$ is correlated with the mean density $\langle \rho \rangle$ of the tidally stripped star, so the orbital period may gradually increase with a decrease in the stellar mass \citep{Rossi2021}. Increased orbital periods and delayed periodic X-ray outbursts have been observed in ESO\,243$-$49 HLX$-$1 \citep{Lin2020}. Additionally, semi-analytic models of a main-sequence star around a Schwarzschild SMBH with $10^7$ M$_{\odot}$ also predicted that the orbital period increases as the star is stripped \citep{Dai2013}. Therefore, a delayed TDE may occur in IGR\,J12580+0134 after 2022. However, because several parameters, such as the post-TDE stellar mass and SMBH spin, are unknown, the exact time of a potential TDE cannot be predicted. 

Motivated by observing a future TDE, we requested \textit{NICER} monitoring observations of IGR\,J12580+0134 at an interval of three days for nearly a year beginning in March 2023. The flexible schedule and superior soft effective area of the \textit{NICER} are practical for identifying potential X-ray outbursts. Although we did not detect a bright X-ray burst similar to that observed in January 2011, we noticed some faint X-ray flares during March--June 2023, which could have been caused by a weakly accreting SMBH, likely indicating the existence of an AGN in NGC\,4845, as previously suggested by \citet{Irwin2015} based on the variable behavior of the source in radio observations. In this paper, we study the \textit{NICER} monitoring observations of IGR\,J12580+0134 recently collected in 2023--2024, along with previous \textit{XMM-Newton} observations taken during the TDE in 2011 and later in 2021. Section~\ref{ngc4845:observation} describes the observations and data reduction. In Section \ref{ngc4845:analyses} we present the results of our timing and principal component analyses, as well as spectral modeling and simulations of X-ray outbursts. We discuss our results in Section~\ref{ngc4845:discussion}, followed by a summary in Section~\ref{ngc4845:summary}.

\section{Observations and Data Reduction}
\label{ngc4845:observation}

IGR\,J12580+0134 was observed using the European Photo Imaging Camera-pn \citep[EPIC-pn;][]{Strueder2001} on board the X-ray Multi-mirror Mission \citep[\textit{XMM-Newton};][]{Jansen2001} on 22 January 2011 (Obs.\,ID 0658400601, PI R.\,Walter) and 15 June 2021 (proposal ID 84417, PI X.\,Shu). In addition, it was recently monitored using the X-ray Timing Instrument (XTI) aboard the Neutron star Interior Composition Explorer \citep[\textit{NICER};][]{Gendreau2016} for nearly a year from 2 March 2023 until 28 February 2024 (proposal ID 6093, PI A.\,Danehkar). The observation log is presented in Table~\ref{ngc4845:obs:log}, which provides the telescope name, the instrument and its configuration, the observation identification number, start and end times (UTC), and exposure time (ks), as well as the source net counts and count rate (count\,s$^{-1}$) of each observation after the subtraction of background counts and the removal of precipitation events. 

\begin{table*}
\begin{center}
\caption[]{Observation log of IGR\,J12580+0134 with \textit{XMM-Newton} and \textit{NICER}.
\label{ngc4845:obs:log}}
\footnotesize
\begin{tabular}{llllccrrr}
\hline\hline
Observatory & Instrument     & Config. & Obs. ID      & \multicolumn{1}{c}{Obs. Start (UTC)}     & \multicolumn{1}{c}{Obs. End (UTC)}        & Exp. (ks)\,$^{\rm \bf a}$  & Count\,$^{\rm \bf a}$      & Cnt.\,Rate\,$^{\rm \bf a}$ \\
\hline
XMM         & EPIC-pn        & Imaging, Medium    & 0658400601   & 2011 Jan 22, 16:46   & 2011 Jan 22, 22:08    &     21.21/10.22 &       55936 &      5.471 \\
XMM         & EPIC-pn        & Imaging, Medium    & 0884370101   & 2021 Jun 15, 21:15   & 2021 Jun 16, 03:34    &      30.20/6.05 &         216 &      0.036 \\
NICER       & XTI            & Photon             & 6593010101   & 2023 Mar 02, 09:48   & 2023 Mar 02, 10:09    &      0.66 &         365 &      0.555 \\
NICER       & XTI            & Photon             & 6593010301   & 2023 Mar 08, 06:44   & 2023 Mar 08, 10:10    &      0.27 &         270 &      1.020 \\
NICER       & XTI            & Photon             & 6593010401   & 2023 Mar 11, 16:49   & 2023 Mar 11, 17:14    &      0.62 &         330 &      0.529 \\
NICER       & XTI            & Photon             & 6593010501   & 2023 Mar 14, 11:32   & 2023 Mar 14, 11:57    &      0.03 &          35 &      1.381 \\
NICER       & XTI            & Photon             & 6593010601   & 2023 Mar 17, 03:10   & 2023 Mar 17, 03:34    &      1.13 &         456 &      0.404 \\
NICER       & XTI            & Photon             & 6593010701   & 2023 Mar 20, 10:12   & 2023 Mar 20, 10:40    &      1.28 &         509 &      0.397 \\
NICER       & XTI            & Photon             & 6593010801   & 2023 Mar 23, 04:38   & 2023 Mar 23, 20:14    &      0.17 &         405 &      2.343 \\
NICER       & XTI            & Photon             & 6593010901   & 2023 Mar 26, 03:51   & 2023 Mar 26, 05:41    &      0.61 &         285 &      0.468 \\
NICER       & XTI            & Photon             & 6593011001   & 2023 Mar 29, 15:40   & 2023 Mar 29, 16:01    &      1.16 &         296 &      0.255 \\
NICER       & XTI            & Photon             & 6593011101   & 2023 Apr 01, 03:52   & 2023 Apr 01, 10:47    &      1.93 &         826 &      0.429 \\
NICER       & XTI            & Photon             & 6593011201   & 2023 Apr 04, 06:14   & 2023 Apr 04, 06:36    &      1.16 &        1137 &      0.983 \\
NICER       & XTI            & Photon             & 6593011301   & 2023 Apr 07, 10:26   & 2023 Apr 07, 10:46    &      0.53 &         335 &      0.629 \\
NICER       & XTI            & Photon             & 6593011401   & 2023 Apr 10, 04:49   & 2023 Apr 10, 05:24    &      1.54 &        2948 &      1.913 \\
NICER       & XTI            & Photon             & 6593011501   & 2023 Apr 13, 04:04   & 2023 Apr 13, 04:29    &      1.34 &         515 &      0.383 \\
NICER       & XTI            & Photon             & 6593011601   & 2023 Apr 16, 01:48   & 2023 Apr 16, 03:35    &      1.54 &         672 &      0.435 \\
NICER       & XTI            & Photon             & 6593011701   & 2023 Apr 19, 02:39   & 2023 Apr 19, 03:00    &      0.75 &         274 &      0.365 \\
NICER       & XTI            & Photon             & 6593011801   & 2023 Apr 22, 00:20   & 2023 Apr 22, 00:41    &      0.38 &         514 &      1.356 \\
NICER       & XTI            & Photon             & 6593011901   & 2023 Apr 25, 09:12   & 2023 Apr 25, 11:03    &      1.42 &         523 &      0.368 \\
NICER       & XTI            & Photon             & 6593012001   & 2023 Apr 28, 03:33   & 2023 Apr 28, 04:05    &      1.13 &        1950 &      1.730 \\
NICER       & XTI            & Photon             & 6593012101   & 2023 May 01, 04:41   & 2023 May 01, 06:01    &      0.66 &         283 &      0.432 \\
NICER       & XTI            & Photon             & 6593012301   & 2023 May 06, 23:49   & 2023 May 07, 01:46    &      1.23 &         612 &      0.497 \\
NICER       & XTI            & Photon             & 6593012401   & 2023 May 10, 00:40   & 2023 May 10, 02:32    &      0.15 &         171 &      1.145 \\
NICER       & XTI            & Photon             & 6593012501   & 2023 May 13, 04:15   & 2023 May 13, 06:06    &      1.70 &         939 &      0.551 \\
NICER       & XTI            & Photon             & 6593012701   & 2023 May 18, 23:59   & 2023 May 19, 09:29    &      0.81 &         195 &      0.240 \\
NICER       & XTI            & Photon             & 6593012801   & 2023 May 22, 00:46   & 2023 May 22, 04:00    &      0.93 &        2030 &      2.180 \\
NICER       & XTI            & Photon             & 6593012901   & 2023 May 25, 16:58   & 2023 May 25, 18:49    &      1.74 &         741 &      0.427 \\
NICER       & XTI            & Photon             & 6593013001   & 2023 May 28, 09:56   & 2023 May 28, 10:19    &      1.05 &         481 &      0.459 \\
NICER       & XTI            & Photon             & 6593013201   & 2023 Jun 03, 03:49   & 2023 Jun 03, 11:32    &      0.37 &         210 &      0.567 \\
NICER       & XTI            & Photon             & 6593013401   & 2023 Jun 09, 11:30   & 2023 Jun 09, 11:50    &      0.65 &         201 &      0.310 \\
NICER       & XTI            & Photon             & 6593013501   & 2023 Jun 12, 16:58   & 2023 Jun 12, 18:50    &      0.34 &         132 &      0.393 \\
NICER       & XTI            & Photon             & 6593013601   & 2023 Jun 15, 06:54   & 2023 Jun 15, 08:42    &      0.35 &         139 &      0.396 \\
NICER       & XTI            & Photon             & 6593013901   & 2023 Jun 24, 00:05   & 2023 Jun 24, 06:24    &      0.50 &         608 &      1.219 \\
NICER       & XTI            & Photon             & 6593014001   & 2023 Jun 28, 06:05   & 2023 Jun 28, 06:26    &      0.90 &         510 &      0.568 \\
NICER       & XTI            & Photon             & 6593014101   & 2023 Jun 30, 02:51   & 2023 Jun 30, 03:15    &      0.62 &         360 &      0.581 \\
NICER       & XTI            & Photon             & 6593014201   & 2023 Jul 03, 02:04   & 2023 Jul 03, 16:24    &      1.02 &         228 &      0.223 \\
NICER       & XTI            & Photon             & 6593014301   & 2023 Jul 06, 05:49   & 2023 Jul 06, 06:15    &      0.43 &         131 &      0.309 \\
NICER       & XTI            & Photon             & 6593014401   & 2023 Jul 09, 03:58   & 2023 Jul 09, 05:47    &      1.04 &         353 &      0.338 \\
NICER       & XTI            & Photon             & 6593015501   & 2023 Aug 11, 18:57   & 2023 Aug 11, 20:46    &      0.05 &         103 &      2.260 \\
NICER       & XTI            & Photon             & 6593016401   & 2023 Dec 14, 15:35   & 2023 Dec 14, 23:33    &      1.02 &         886 &      0.867 \\
NICER       & XTI            & Photon             & 6593016501   & 2023 Dec 17, 02:26   & 2023 Dec 17, 04:18    &      1.63 &        1171 &      0.719 \\
NICER       & XTI            & Photon             & 6593016601   & 2023 Dec 20, 10:56   & 2023 Dec 20, 12:49    &      1.14 &         496 &      0.435 \\
NICER       & XTI            & Photon             & 6593016701   & 2023 Dec 22, 09:24   & 2023 Dec 22, 12:38    &      0.96 &         309 &      0.321 \\
NICER       & XTI            & Photon             & 6593017401   & 2024 Jan 13, 01:43   & 2024 Jan 13, 02:07    &      0.81 &         175 &      0.216 \\
NICER       & XTI            & Photon             & 6593017501   & 2024 Jan 16, 00:58   & 2024 Jan 16, 23:22    &      0.74 &         370 &      0.498 \\
NICER       & XTI            & Photon             & 6593017601   & 2024 Jan 19, 17:16   & 2024 Jan 19, 17:38    &      1.02 &         493 &      0.485 \\
NICER       & XTI            & Photon             & 6593017701   & 2024 Jan 22, 11:59   & 2024 Jan 22, 13:47    &      1.50 &         452 &      0.301 \\
NICER       & XTI            & Photon             & 6593017801   & 2024 Jan 25, 03:29   & 2024 Jan 25, 05:17    &      0.38 &         149 &      0.389 \\
NICER       & XTI            & Photon             & 6593017901   & 2024 Jan 27, 21:59   & 2024 Jan 27, 23:46    &      0.73 &         237 &      0.322 \\
NICER       & XTI            & Photon             & 6593018101   & 2024 Feb 03, 04:20   & 2024 Feb 03, 07:37    &      0.41 &         147 &      0.358 \\
NICER       & XTI            & Photon             & 6593018201   & 2024 Feb 06, 08:16   & 2024 Feb 06, 13:10    &      0.25 &         157 &      0.630 \\
NICER       & XTI            & Photon             & 6593018301   & 2024 Feb 09, 04:27   & 2024 Feb 09, 06:16    &      0.77 &         270 &      0.349 \\
NICER       & XTI            & Photon             & 6593018501   & 2024 Feb 16, 00:57   & 2024 Feb 16, 04:17    &      1.43 &         336 &      0.235 \\
NICER       & XTI            & Photon             & 6593018601   & 2024 Feb 18, 10:05   & 2024 Feb 18, 16:25    &      1.12 &         277 &      0.248 \\
NICER       & XTI            & Photon             & 6593018701   & 2024 Feb 21, 12:35   & 2024 Feb 21, 14:22    &      1.10 &         307 &      0.278 \\
NICER       & XTI            & Photon             & 6593018801   & 2024 Feb 24, 05:32   & 2024 Feb 24, 05:52    &      0.85 &         197 &      0.230 \\
NICER       & XTI            & Photon             & 6593018901   & 2024 Feb 28, 17:53   & 2024 Feb 28, 18:15    &      1.09 &         314 &      0.287 \\
\hline
\end{tabular}
\end{center}
\begin{tablenotes}
\footnotesize
\item[1]\textbf{Note.} $^{\rm \bf a}$ Source net counts and rates over the energy range of 0.3--10\,keV after background subtraction and precipitating electron removal. The net exposure times (after the slash in XMM), counts, and count rates are associated with events filtered from non-X-ray flares caused by geomagnetic electrons in the outer magnetosphere.
\end{tablenotes}
\end{table*}

We acquired the original data files (ODFs) from the XMM-Newton Science Archive\footnote{\href{https://nxsa.esac.esa.int/}{https://nxsa.esac.esa.int/}} and reprocessed them using the XMM Science Analysis Software \citep[\textsc{sas} v\,20.0.0;][]{Gabriel2004} and calibration files (XMMCCF-REL-391) released on 25 October 2022. The \textsc{sas} tasks \textsf{cifbuild} and \textsf{odfingest} were employed to generate calibration index files (CIF) and reprocess ODFs, respectively. The pn events were created using the \textsc{sas} tasks \textsf{epproc}. To address non-X-ray flares caused by geomagnetic electrons, we eliminated events exceeding the count rates of 0.4 c\,s$^{-1}$ detected using the single-pixel (PATTERN\,$=$\,0) light curves binned at 100\,s in the energy band of 10--12\,keV. The flare-cleaned events were further filtered using the \textsc{sas} tool \textsf{evselect} to retain only patterned events (PATTERN\,$\leqslant$\,4) and appropriate PI channels (200\,$<$\,PI\,$<$\,15000) and to exclude defective pixels (FLAG\,$=$\,0). Source spectra, redistribution matrix files (RMF), and auxiliary response files (ARF) were produced using the \textsc{sas} task \textsf{especget} from a circular aperture with a $36''$ radius centered on the brightest source. Background spectra were created from an equally sized circular region on the same chip, positioned away from the source. 

The \textit{NICER} data were obtained from the High Energy Astrophysics Science Archive Research Center (HEASARC)\footnote{\href{https://heasarc.gsfc.nasa.gov/}{https://heasarc.gsfc.nasa.gov/}} and calibrated using the \textsc{nicerdas} software (v\,2024-02-09\_V012A) as part of HEAsoft \citep[v\,6.33.2;][]{NASAHEASARC2014} and the current calibration files released on 06 February 2024. The \textsf{niprelflarecalc} procedure was used to add additional columns for the precipitating electron (PREL) flaring index to the ``MKF'' filter file for each observation. The task \textsf{niprelflaregti} was then applied to the MKF file to construct flare-cleaned Good Time Interval (GTI) tables from the PREL index. We applied the flare-cleaned GTI file to Level-2 standard calibration and filtering processes for each observation using the task \textsf{nicerl2}. To exclude the charged particle background, we filtered out any burst-like behaviors by restricting events to elevation angles of $\geqslant30^{\circ}$ above the Earth's limb (\textsf{elv=30}), offsets of $\geqslant40^{\circ}$ from the bright Earth (\textsf{br\_earth=40}), the default overshoot range of $\leqslant30$\,count\,s$^{-1}$, 
and the magnetic cut-off rigidity (\textsf{COR\_SAX}) of $\geqslant 4$\,GeV/c (\textsf{cor\_range=``4-*''}) via the task \textsf{nicerl2}. The cleaned event files were further utilized to create pulse height amplitude (PHA) spectra, backgrounds, and redistribution and response files (RMF and ARF) using the procedure \textsf{nicerl3-spect}, to which we again supplied the non-flare GTI tables. Using the FTOOL task \textsc{ftgrouppha}, each spectrum was also binned with the optimal binning approach \citep{Kaastra2016}, with each bin containing at least 30 counts. The background  of each dataset was estimated using the \textsf{SCORPEON} model (v\,23), which was designed to handle PREL flares, and was saved in the PHA file format.
The source net counts and count rates (counts\,s$^{-1}$) presented in Table~\ref{ngc4845:obs:log} are associated with the \textit{NICER} events cleaned from non-X-ray flares caused by precipitating electrons in the outer magnetosphere. 
The background counts estimated using the \textsf{SCORPEON} model were also subtracted from the net counts and count rates of each dataset.


\section{Data Analyses and Results}
\label{ngc4845:analyses}

\subsection{Timing Analysis}
\label{ngc4845:time}

To interpret the evolution of the previous TDE and identify a potential TDE, we conducted timing analysis using the light curves of the \textit{XMM-Newton}/EPIC-pn and \textit{NICER} observations in three different energy bands: soft ($S$: 0.4--1.1\,keV), medium ($M$: 1.1--2.6\,keV), and hard ($H$: 2.6--10\,keV). We generated \textit{XMM-Newton} light curves of the source and background at an interval step of 600\,s for the aforementioned energy ranges using the \textsc{sas} program \textsf{evselect}. The \textit{NICER} light curves of the source and background were produced using the \textsc{nicerdas} task \textsf{nicerl3-lc}, adopting the \textsf{SCORPEON} background model. \textit{NICER} light curves were binned at 120\,s, providing sufficient temporal resolution to identify spectral changes such as coronal flares occurring on timescales of a few minutes.  To account for the gradual decrease in instrument sensitivity, we also adjusted the second \textit{XMM-Newton} light curves upward relative to the first observation based on the integration of the ARF effective areas across the specified energy range in a manner similar to that implemented in \citet{Danehkar2024}.
 
\begin{figure*}
\begin{center}
\includegraphics[width=0.48\textwidth, trim = 0 0 0 0, clip, angle=0]{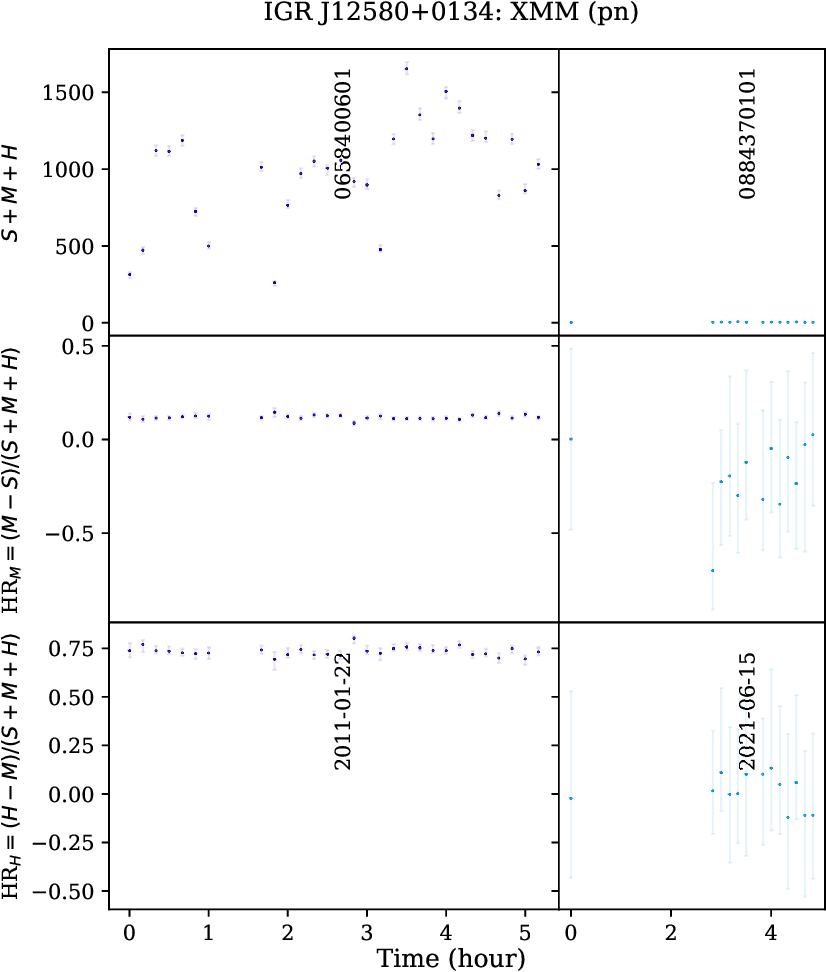}%
\includegraphics[width=0.48\textwidth, trim = 0 0 0 0, clip, angle=0]{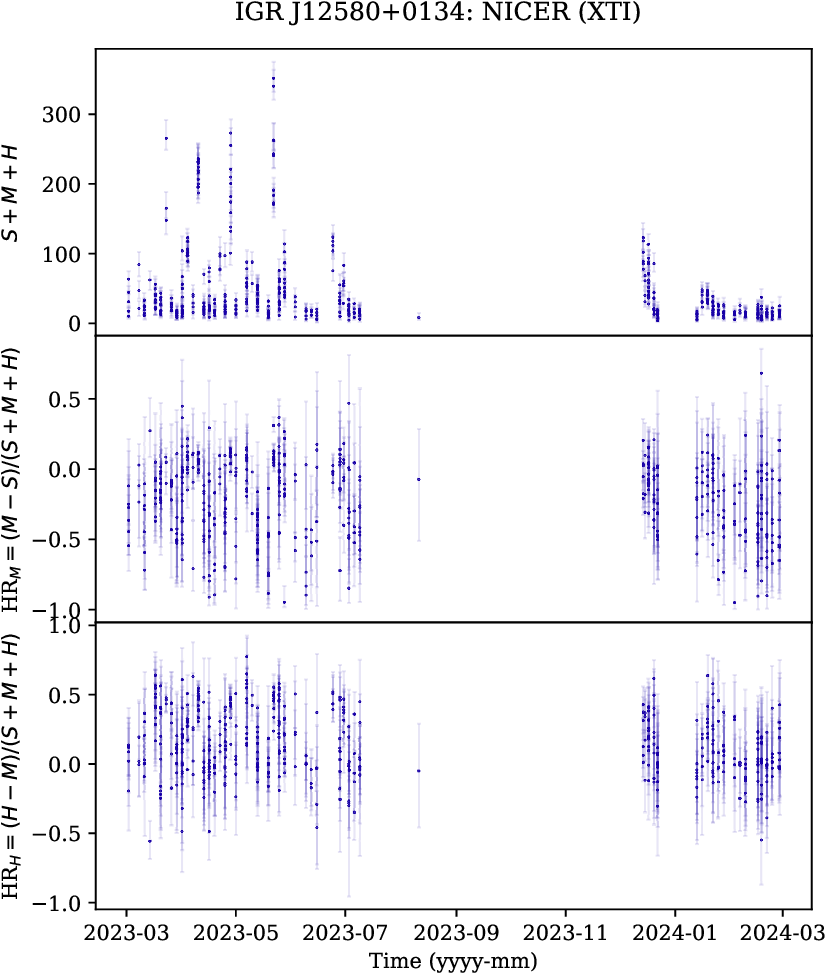}%
\end{center}
\caption{The light curves of IGR\,J12580+0134 in the broad bands ($S+M+H$; in counts), together with the corresponding hardness ratios ${\rm HR}_{M} = (M-S)/(S+M+H)$ and ${\rm HR}_{H} = (H-M)/(S+M+H)$ binned at 600 sec and 120 sec made with the \textit{XMM-Newton} and \textit{NICER} observations, respectively.
\label{ngc4845:fig:lc}
}
\end{figure*}

To detect any temporary transitions in the X-ray source, we calculated the following hardness ratios using the light curves from the soft ($S$), medium ($M$), and hard ($H$) bands:  
\begin{align}
& \mathrm{HR}_{M} = \frac{M-S}{S+M+H}, \quad \mathrm{HR}_{H} = \frac{H-M}{S+M+H}. \label{eq_1}
\end{align}
Hardness ratios can be employed to examine X-ray variability in diverse X-ray sources, such as AGN, X-ray binaries and symbiotic stars \citep[e.g.,][]{Prestwich2003,Sreehari2021,Danehkar2021}. In particular, the hardness ratio ${\rm HR}_{H}$ allows us to investigate changes in the power-law continuum, which typically originates from the corona in radio-quiet AGN. The hardness ratio ${\rm HR}_{M}$ also exhibits variation in the medium band, which is often associated with warm absorbers in AGN \citep[e.g.,][]{Reynolds1997,Kaspi2002}.

Figure~\ref{ngc4845:fig:lc} shows the broadband ($S+M+H$) time series of IGR\,J12580+0134, which was adjusted for the background using the Bayesian Estimator for Hardness Ratios \citep[BEHR;][]{Park2006} on the source and background light curves. The figure also shows the corresponding hardness ratios defined by Eq. (\ref{eq_1}). The light-curve and hardness ratio uncertainties were determined using the Gibbs sampler based on the Monte Carlo simulations. While the \textit{XMM-Newton} light curves show random, intense variations in January 2011, the source was in an exceptionally bright state. However, it was extremely faint in the \textit{XMM-Newton} observation collected in June 2021. We also noticed some dramatic increases in X-ray brightness in the \textit{NICER} observations from March to June 2023, followed by a fainter state in 2024. The source appeared to be significantly variable in January 2011 and March--June 2023. We caution that these two instruments have different effective areas, so their counts are not comparable. Nevertheless, the X-ray fluxes derived from our spectral analysis in \S\,\ref{ngc4845:spec} imply that the X-ray spectrum recorded by the \textit{XMM-Newton} EPIC-pn during the TDE was about 120 times brighter than the time-averaged \textit{NICER} spectrum. Moreover, the hardness ratio ${\rm HR}_{H}$ was unusually high during the TDE in January 2011, although we did not observe such a situation during March--June 2023. Moreover, the hardness ratio ${\rm HR}_{M}$ was slightly elevated in 2011, likely indicating accretion onto the SMBH. However, both hardness ratios remained the same over the course of the \textit{NICER} monitoring observations over 2023--2024.
\begin{figure*}
\begin{center}
\includegraphics[height=0.53\textheight, trim = 0 0 0 0, clip, angle=0]{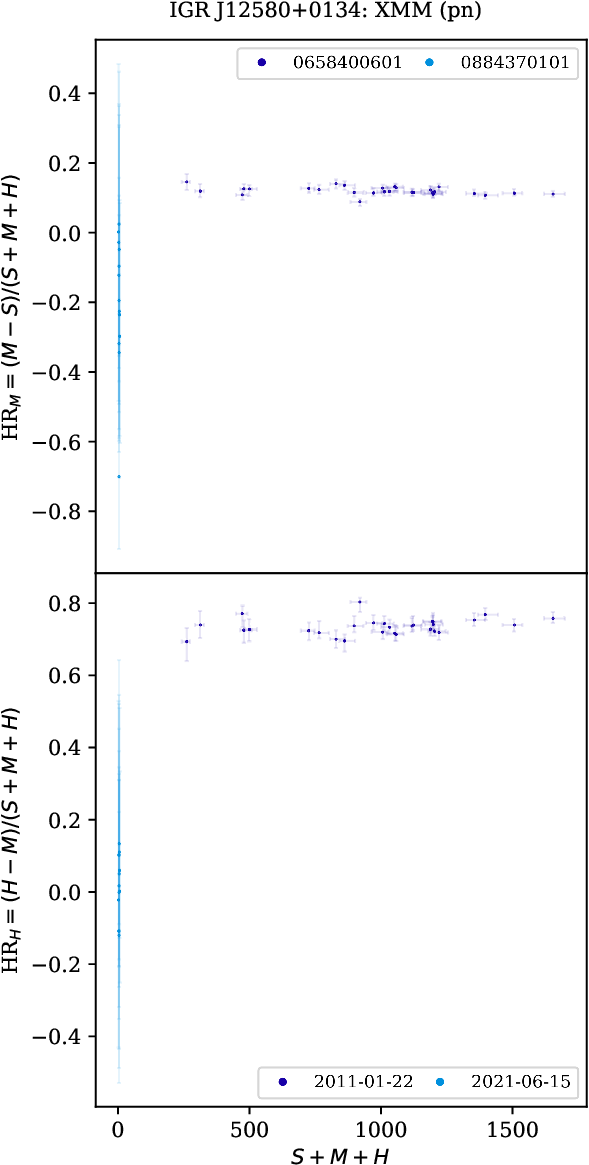}%
\includegraphics[height=0.53\textheight, trim = 0 0 0 0, clip, angle=0]{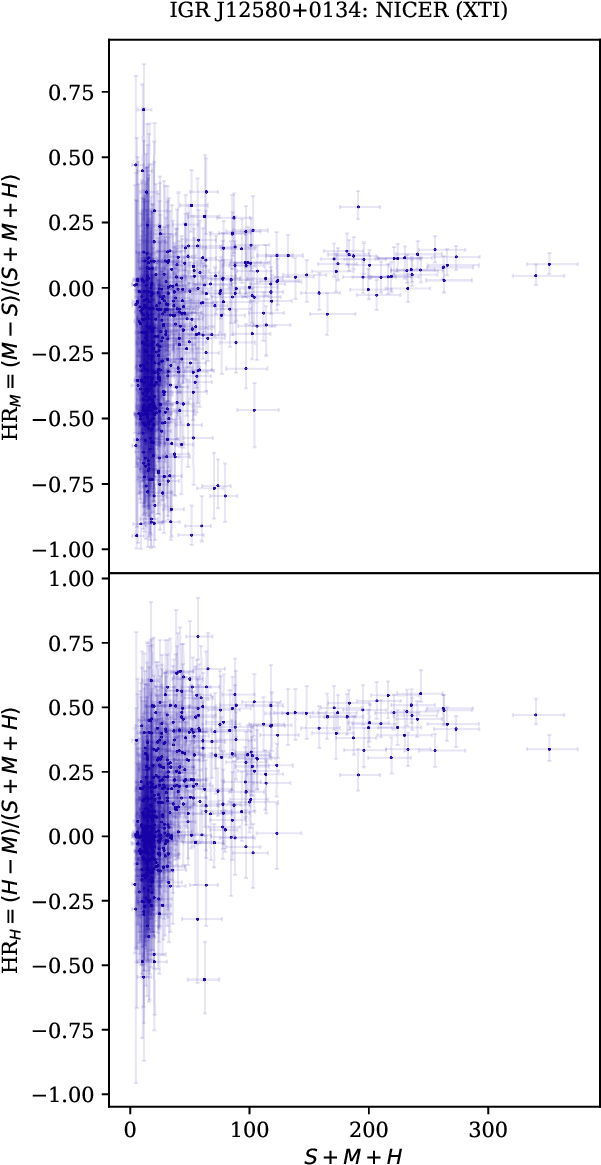}%
\end{center}
\caption{The hardness ratio diagrams of IGR\,J12580+0134: the hardness ratios ${\rm HR}_{M} = (M-S)/(S+M+H)$ and ${\rm HR}_{H} = (H-M)/(S+M+H)$ plotted against the broad band ($S+M+H$; in counts) using the light curves binned at 600 sec and 120 sec from the \textit{XMM-Newton} and \textit{NICER} observations, respectively.
\label{ngc4845:fig:hdr}
}
\end{figure*}

\begin{figure*}
\begin{center}
\includegraphics[width=0.95\textwidth, trim = 0 0 0 0, clip, angle=0]{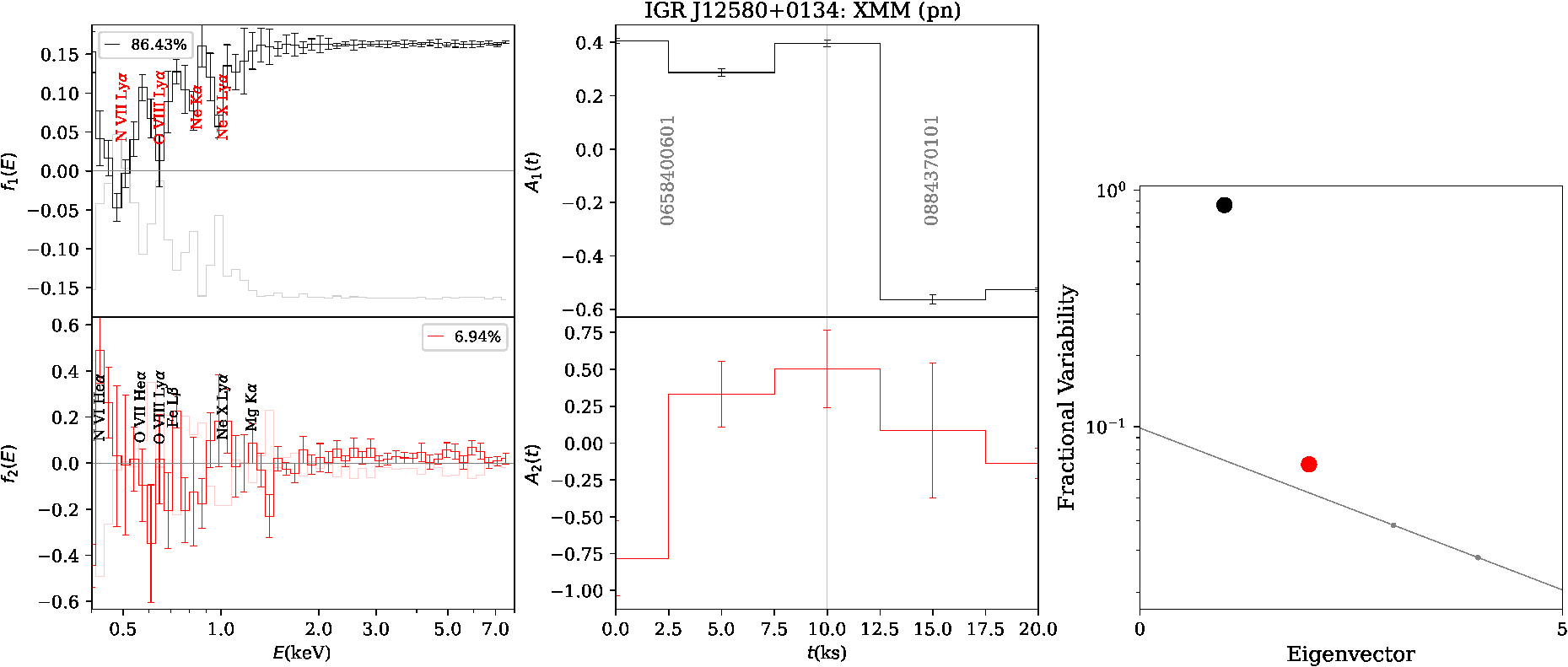}\\
\includegraphics[width=0.95\textwidth, trim = 0 0 0 0, clip, angle=0]{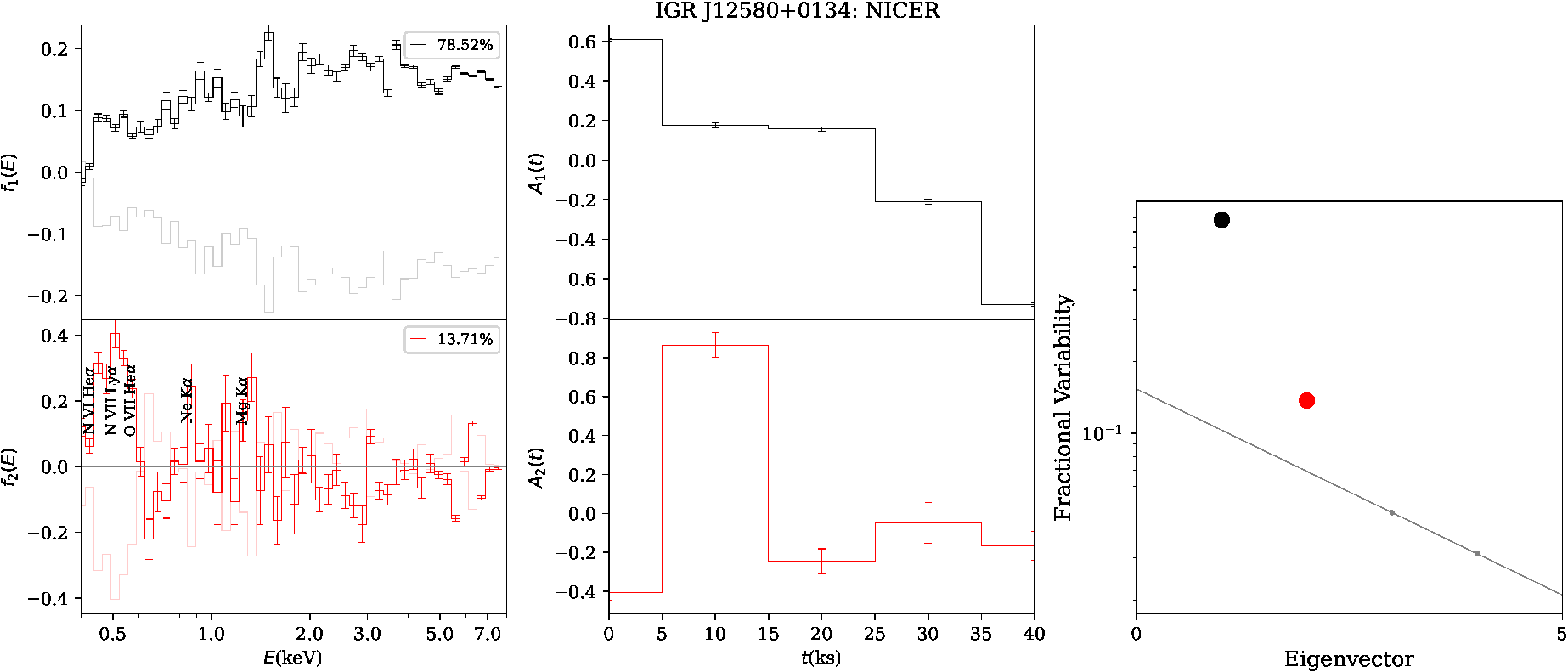}%
\end{center}
\caption{The normalized PCA spectra $f_{i}(E)$ (left) and the corresponding light curves $A_{i}(t)$ (middle) obtained from the \textit{XMM-Newton}/EPIC-pn (top panels) and \textit{NICER} observations (bottom panels) of IGR\,J12580+0134, along with the LEV diagrams (right), each showing the linear regression among normalized logarithmic eigenvalues and eigenvector orders higher than 2.
The energy levels, where atomic spectral features may be present, are labeled in the spectra. The number of eigenvalues in each LEV diagram corresponds to the number of time-sliced spectra used in PCA.
\label{ngc4845:fig:pca:1}
}
\end{figure*}

The hardness ratio diagrams shown in Fig.\,\ref{ngc4845:fig:hdr} were created by plotting the previously mentioned hardness ratios against broadband light curves, helping to identify potential TDE or X-ray flares from the corona. The X-ray source demonstrated a transient effect in the \textit{XMM-Newton} observation in January 2011 that was associated with the TDE. We see that the hardness ratios ${\rm HR}_{M}$ and ${\rm HR}_{H}$ were about 0.15 and 0.75 during the TDE, respectively. Similarly, it seems that there are some transient effects over the course of the \textit{NICER} observations from March to June 2023 while also exhibiting small-scale random variations. Again, the hardness ratio ${\rm HR}_{M}$ was approximately 0.15 during the transient events in 2023, similar to the TDE in 2011, whereas the hardness ratio ${\rm HR}_{H}$ was approximately 0.5 during the X-ray flares, which was lower than that in the TDE in 2011.  The source was much fainter during the transient events in 2023 than in the TDE in 2011, implying the presence of X-ray flares rather than a TDE. These flares could indicate a weakly accreting SMBH.

\subsection{Principal Component Analysis}
\label{ngc4845:pca}

To better understand the X-ray brightness changes during the TDE and X-ray flares, we employed PCA to separate the spectral components primarily responsible for the X-ray variations in this object. PCA, also known as the ``Hotelling transform'' \citep{Hotelling1933}, is a highly effective technique for deconstructing spectral components in variable spectra and has frequently been used in the field of statistics and machine learning in astronomy \citep[see, e.g.,][]{Wall2012,Ivezic2020}, such as for gaining insights into the spectra of stars \citep{Deeming1964,Whitney1983} and galaxies \citep{Faber1973,Bujarrabal1981,Efstathiou1984,Mittaz1990}. This approach has also been employed to analyze the X-ray variability of AGN \citep[e.g.,][]{Vaughan2004,Miller2008,Parker2014,Parker2017,Parker2018}, X-ray binaries \citep{Koljonen2015,Koljonen2013}, symbiotic stars \citep{Danehkar2024}, and starburst regions \citep{Danehkar2024a}.  

In our PCA, we utilized a modified version of the Python program \textsc{pca}\footnote{\href{https://github.com/xstarkit/pca-code}{https://github.com/xstarkit/pca-code}} developed by \citet{Parker2015,Parker2017}, which is based on the singular value decomposition \citep[SVD][]{Press1997} function from the Python package \textsf{NumPy} \citep{Harris2020}. PCA involves breaking down a rectangular (${m \times n}$) matrix $\mathbf{M}$ containing $n$ time-order spectra with $m$ energy bins into 
$\mathbf{M} = \mathbf{U}\mathbf{\Sigma}\mathbf{V}^{\intercal}$, where the square matrix $\mathbf{U}$ of order $m$ includes principal components (spectra $f_i(E)$ in our case), the rectangular ($m \times n$) diagonal matrix $\mathbf{\Sigma}$ contains square roots of eigenvalues ($\mathbf{\Lambda}=\mathbf{\Sigma}^{\intercal}\mathbf{\Sigma}=\mathrm{diag}(\lambda_1,\ldots,\lambda_n)$; contribution fractions in variability), and the square matrix $\mathbf{V}$ of order $n$ comprises eigenvectors \citep[light curves $A_i(t)$ here;][]{Parker2017}. This program transfers the provided time-segmented spectra into a set of energy-binned spectral data, which are then decomposed using the SVD function into principal spectral components $f_i(E)$ and their corresponding time-binned light curves $A_i(t)$, along with the component contribution percentages calculated using the corresponding normalized eigenvalues $\lambda_i$. Error calculations involve randomly perturbing the given spectra and recalculating the SVD \citep[refer to][]{Miller2007}.

To perform PCA, we filtered the events of the \textit{XMM-Newton} observations to generate time-sliced spectra at 5\,ks intervals according to the data reduction method described in detail by \citet{Danehkar2024}. To produce the time-segmented \textit{NICER} data, we created a series of lists, each containing a group of datasets with a total exposure of approximately 10\,ks. The datasets of each list were then merged using the \textsf{niobsmerge} task, which was applied to Level-2 products generated by the \textsf{nicerl2} task described in \S\,\ref{ngc4845:observation}. Similar to standard \textit{NICER} data reduction, the corresponding PREL indices and GTI tables were produced using the tasks \textsf{niprelflarecalc} and \textsf{niprelflaregti}, respectively. Time-sliced spectra of the source and background were then generated using \textsf{nicerl3-spect}, in which the \textsf{SCORPEON} background model was assigned, and the GTI tables were also applied.  We chose intervals of 5\,ks (XMM) and 10\,ks (NICER) to provide sufficient counts in each time-segmented spectrum for PCA as well as sufficient time resolution for disclosing variations occurring on time scales of about 3 hours. After data reduction, we supplied the time-sliced spectra of the source and background, as well as the ARF, to the Python \textsc{pca} code to generate the principal spectra (PCA components), light curves (eigenvectors), and fractional contributions (normalized eigenvalues). We employed the effective area column extracted from the ARF to transform the count value in each energy bin into a relative flux value. Background-corrected PCA spectra were then used to create a mean spectrum, which was then used to normalize the PCA spectra.  

The spectra $f_i(E)$ and light curves $A_i(t)$ of the first two PCA components ($i=1,2$) derived from the \textit{XMM-Newton}/EPIC-pn and \textit{NICER} observations are shown in Fig.\,\ref{ngc4845:fig:pca:1}, along with their respective log-eigenvalue (LEV) diagrams. The LEV diagrams plot the normalized eigenvalues of the PCA components against the corresponding eigenvector orders. The number of eigenvalues is associated with the number of time-sliced spectra used in PCA. Each of these diagrams shows a line representing the linear regression between the logarithmic values of the normalized eigenvalues and eigenvector orders, starting from the third order, which helps differentiate the statistically significant PCA components from those potentially caused by background noise \citep[see, e.g.,][]{Koljonen2013,Parker2017}. Based on the LEV diagrams, the first two components appear to be statistically significant for producing the TDE in the \textit{XMM-Newton} data and X-ray flares in the \textit{NICER} monitoring observations.

The first component of the \textit{XMM-Newton} data resembles the spectral characteristics of an absorbed power-law component above 1 keV, aligned with the model employed by \citet{Nikolajuk2013}, who also included unabsorbed blackbody emission below 1 keV.  Similarly, the first component of the \textit{NICER} observations exhibits power-law-like features, albeit with a significantly lower absorbing density column of the foreground material as found by our spectral analysis in the next subsection (\S\,\ref{ngc4845:spec}). The normalized eigenvalues revealed variability fractions of approximately 86\% and 79\% for the first components in the \textit{XMM-Newton} and \textit{NICER} data, respectively. This suggests that increases in the source continuum are primarily responsible for the horizontal pattern observed in the hardness ratio diagrams (Fig.\,\ref{ngc4845:fig:hdr}). The primary principle component resulting from the \textit{XMM-Newton} data corresponds to the TDE in January 2011, which was attributed to a significant increase in source brightness. However, the main principle component from the \textit{NICER} monitoring observations is associated with X-ray flares occurring from March to June 2023, as seen in Fig.\,\ref{ngc4845:fig:lc}. The light curves of the first components demonstrate that as the source experienced brightening events, the continuum levels of the power-law spectrum were elevated. The second-order PCA components from both the \textit{XMM-Newton} and \textit{NICER} data, with small variability fractions of approximately 7\% and 14\%, respectively, depict time series that have time-lagged peaks relative to the light curves of the first-order PCA components. 

Some spectral features in the soft excess ($< 2$\,keV) are more pronounced than those in the hard excess. The valley features in the first-order PCA spectrum of the \textit{XMM-Newton} data could be associated with absorption lines produced by the line-of-sight neutral and ionized material, such as \ionic{N}{vii} Ly$\alpha$ 0.5\,keV, \ionic{O}{viii} Ly$\alpha$ 0.65\,keV, and Ne K$\alpha$ 0.85\,keV. However, these features are absent from the first-order PCA spectrum of the \textit{NICER} data. In particular, our spectral analysis in the following subsection (\S\,\ref{ngc4845:spec}) indicates an extremely low absorbing column of the line-of-sight material during the \textit{NICER} monitoring observations. The peak features in the second-order PCA spectra could be related to emission lines \citep[see, e.g.,][]{Danehkar2024,Danehkar2024a}. These features may correspond to H-like and He-like emission lines, such as \ionic{N}{vi} He$\alpha$ 0.43\,keV and \ionic{O}{vii} He$\alpha$ 0.57\,keV. The second-order components suggest that there could be some emission lines in the soft excess, likely from thermally ionized or shock-excited  winds/jets, which can be modeled by a soft thermal plasma emission component, such as \textsc{apec}, as performed by \citet{Perlman2022}. 

The second-order PCA components appear to stem from phenomena that lag behind the peak of the TDE in \textit{XMM-Newton} or X-ray flares in \textit{NICER}. This suggests that they may be associated with the collision of winds or jets resulting from the TDE or flares with the surrounding medium away from the SMBH following the beginning of the TDE or X-ray flares. However, we caution that PCA, which separates uncorrelated eigenvectors of the matrix containing time-order data, may not inherently preserve physical causality. PCA decomposition imposes principal components to be statistically uncorrelated, so spurious lags may arise from it, as PCA assigns part of the variability to a new component rather than fully integrating it into the earlier component.

\begin{figure}
\begin{center}
\includegraphics[width=0.443\textwidth, trim = 0 0 0 0, clip, angle=0]{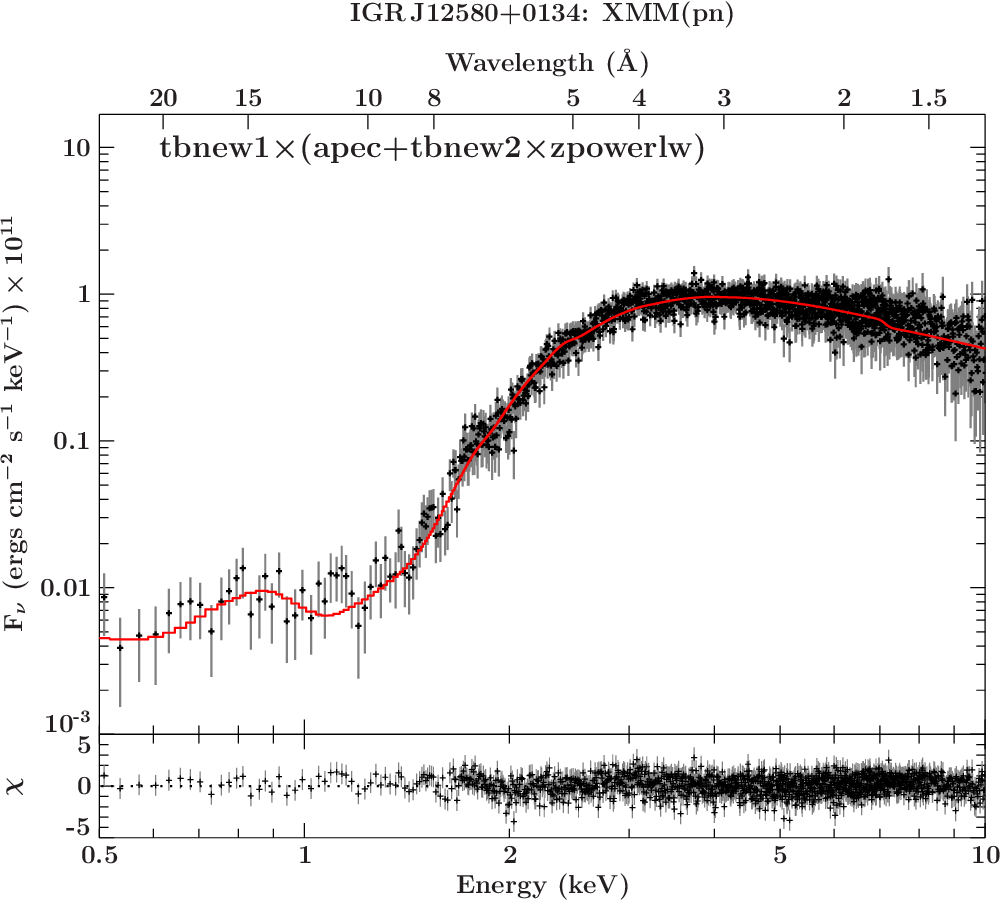}\\
\includegraphics[width=0.443\textwidth, trim = 0 0 0 0, clip, angle=0]{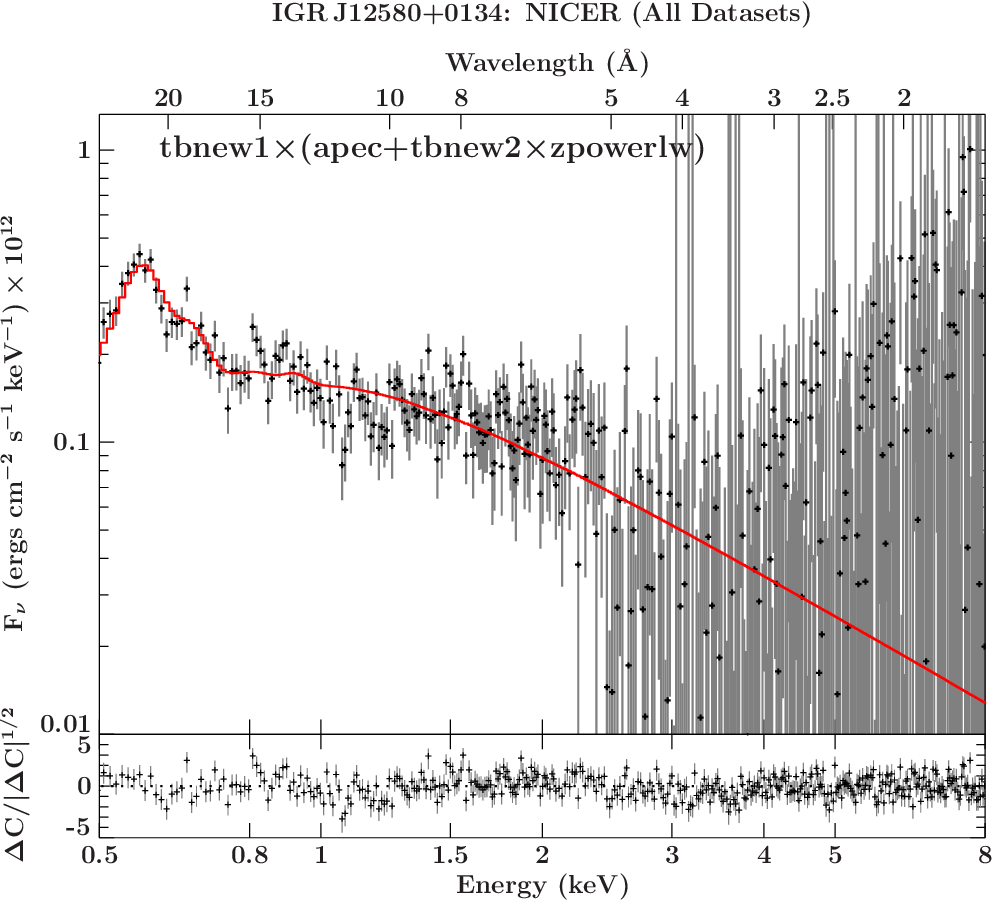}\\
\includegraphics[width=0.443\textwidth, trim = 0 0 0 0, clip, angle=0]{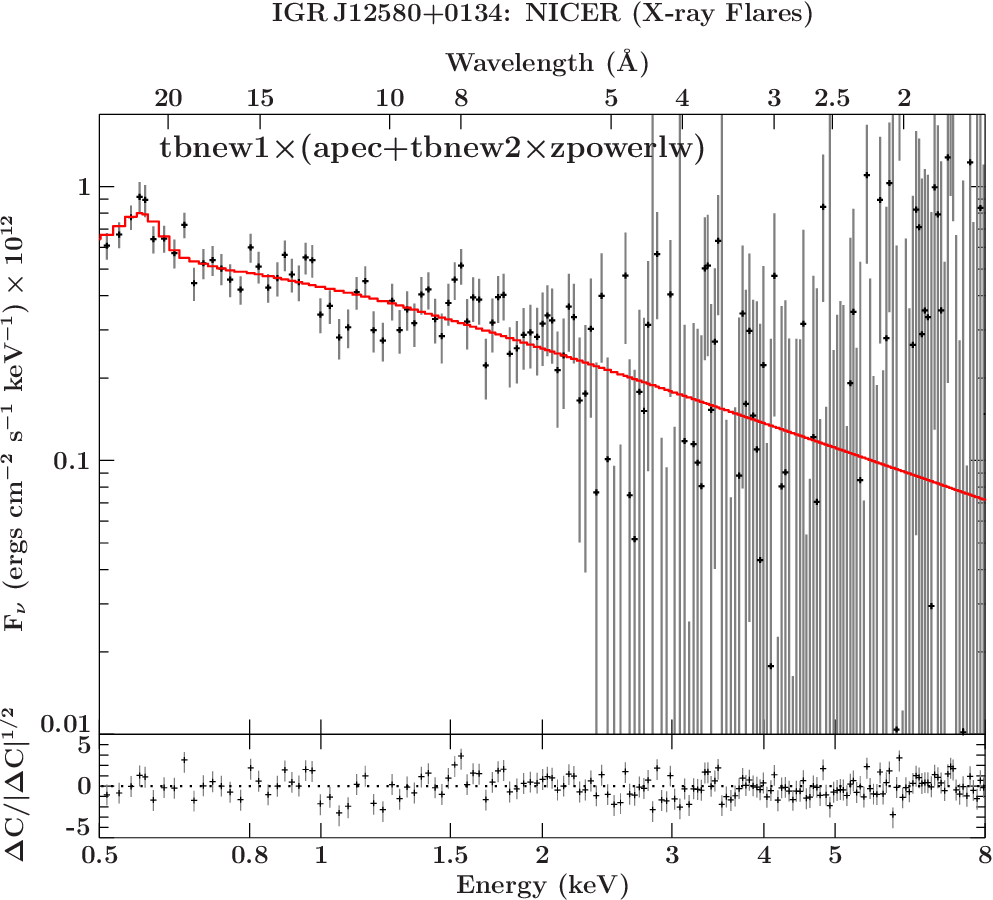}%
\end{center}
\caption{The \textit{XMM-Newton} spectrum during the TDE (top; January 2011), \textit{NICER} time-averaged spectrum of all monitoring observations (middle; 2023--2024) and time-averaged spectrum during X-ray flares (bottom; March--June 2023) of IGR\,J12580+0134, fitted to the spectral model \textsf{tbnew1}\,$\times$\,(\textsf{apec}\,$+$\,\textsf{tbnew2}\,$\times$\,\textsf{zpowerlw}) plotted by red color lines. 
\label{ngc4845:fig:spec}
}
\end{figure}


\subsection{Spectral Analysis}
\label{ngc4845:spec}

To further assess the X-ray properties of IGR\,J12580+0134, we conducted spectral analyses of the \textit{XMM-Newton}/EPIC-pn spectrum and time-averaged \textit{NICER} observations. We grouped the NICER observations into two sets: all datasets and those with X-ray flares. The seven datasets\footnote{Obs. IDs 659301-0801, -1201, -1401, -1801, -2001, -2801, and -3901.} with count rates higher than 0.9\, count\,s$^{-1}$ and total counts higher than 400 in Table~\ref{ngc4845:obs:log}  were classified as those containing X-ray flares, which manifest brightening features seen in the broad bands ($S+M+H$) during March--June 2023, as seen in Fig.\,\ref{ngc4845:fig:lc}.

Our spectral modeling was implemented using the MIT Chandra X-ray Center's Interactive Spectral Interpretation System\footnote{\href{https://space.mit.edu/asc/isis/}{https://space.mit.edu/asc/isis/}} \citep[\textsc{isis} v.\,1.6.2-51;][]{Houck2000}, which is designed to model high-resolution X-ray spectra and provides access to the spectral models of the X-ray spectral fitting package \textsc{xspec} \citep[v.\,12.14.0;][]{Arnaud1996}. The \textit{XMM-Newton}/EPIC-pn spectrum was binned to ensure a minimum of 15 counts per energy bin. Spectral modeling of the \textit{XMM-Newton} spectrum was performed using the chi-square ($\chi^2$) statistic method together with the Levenberg--Marquardt optimization method \citep[\textsf{mpfit};][]{More1978,More1980}.\footnote{\textsc{isis} uses the C function \textsf{mpfit} translated from Fortran by S.~L.~Moshier and enhanced by \citet{Markwardt2009}.} All the \textit{NICER} datasets and those with X-ray flares were merged using the \textsc{nicerdas} task \textsf{niobsmerge}, and the corresponding spectra were created with \textsf{nicerl3-spect} after PREL non-X-ray flare removal. The \textit{NICER} spectra were also rebinned using the \textsc{isis} function \textsf{rebin\_data} to contain a minimum of 60 counts per bin. As the \textit{NICER} spectra have low source counts and intense backgrounds, we simultaneously fitted the source and background using the ``Cash'' statistic maximum-likelihood function \citep[C-stat;][]{Cash1979} in conjunction with the \textsf{subplex} optimization method \citep[][]{Rowan1990}.\footnote{\textsc{isis} utilizes the \textsf{subplex} library from \href{https://www.netlib.org/opt/}{https://www.netlib.org/opt/}.} The background spectra were modeled using the sum of the two broken power-law components. Our spectral modeling was restricted to an energy range of 0.4--10\,keV.

To model the spectra, we used an absorbed power law (\textsf{zpowerlw}) and an optically thin plasma model for collisional ionization from the Astrophysical Plasma Emission Code \citep[\textsc{apec};][]{Smith2001}. The absorption component was created using a \textsf{tbnew} component \citep{Wilms2000}. To account for foreground Galactic absorption, we also included another \textsf{tbnew} component, whose column density was fixed at the total Galactic mean column density of hydrogen ($N_{\rm H,tot}= 1.56\times 10^{20}$\,cm$^{-2}$) from the UK Swift Science Data Centre\footnote{\href{https://www.swift.ac.uk/analysis/nhtot/}{https://www.swift.ac.uk/analysis/nhtot/}} \citep[UKSSDC;][]{Willingale2013}. We assumed the interstellar medium (ISM) composition of \citet{Wilms2000} to be solar abundances. Our spectral model is expressed as \textsf{tbnew1}\,$\times$\,(\textsf{apec}\,$+$\,\textsf{tbnew2}\,$\times$\,\textsf{zpowerlw}). We set the redshift parameter in the components \textsf{apec}, \textsf{tbnew2}, and \textsf{zpowerlw} to the rest frame of NGC\,4845 \citep[$z=0.003589$;][]{Yu2022}. The source spectra and corresponding best-matched spectral models (red lines) are shown in Fig.~\ref{ngc4845:fig:spec}.

\begin{table*}
\begin{center}
\caption{Best-fitting parameters for the spectral models of the \textit{XMM-Newton} and \textit{NICER} observations of IGR\,J12580+0134.  
\label{ngc4845:model:param}
}
\begin{tabular}{llccc}
\hline\hline
\noalign{\smallskip}
{Component} & {Parameter} & {\textit{XMM-Newton}}   & {\textit{NICER}}  & {\textit{NICER}} \\
                 &  & {(TDE 2011)}   & {(All Datasets)}  & {(X-ray Flares)} \\
\noalign{\smallskip}
\hline\noalign{\smallskip} 
 & \multicolumn{4}{c}{Model: \textsf{tbnew1}\,$\times$\,(\textsf{apec}\,$+$\,\textsf{tbnew2}\,$\times$\,\textsf{zpowerlw})} \\
\noalign{\smallskip}
\hline\noalign{\smallskip}
\textsf{tbnew1}\,$^{\rm \bf a}$ & $N_{\rm H,Gal}$ ($10^{20}$\,cm$^{-2}$) \dotfill & $ { 1.56 } $ &  $ { 1.56 } $ &  $ { 1.56 } $ \\
\noalign{\smallskip} 
\textsf{apec}\,$^{\rm \bf b}$ & $K_{\rm ap}$ ($10^{-5}$\,cm$^{-5}$) \dotfill & $ {  1.41 }_{ -0.38 }^{ + 0.45 } $ &  $ {  8.77 }_{ -0.93 }^{ + 0.93 } $ &  $ { 27.96 }_{ -22.16 }^{ +22.16 } $ \\
              & $kT$ (keV) \dotfill & $ {  0.78 }_{ -0.19 }^{ + 0.20 } $ &  $ {  0.17 }_{ -0.01 }^{ + 0.01 } $ &  $ {  0.11 }_{ -0.01 }^{ + 0.02 } $ \\
\noalign{\smallskip} 
\textsf{tbnew2} & $N_{\rm H,pl}$ ($10^{22}$\,cm$^{-2}$) \dotfill & $ { 11.22 }_{ -0.25 }^{ + 0.25 } $ &  $ {  0.30 }_{ -0.06 }^{ + 0.07 } $ &  $ {  0.08 }_{ -0.06 }^{ + 0.06 } $ \\
\noalign{\smallskip} 
\textsf{zpowerlw} & $K_{\rm pl}$ ($10^{-4}$\,$\frac{\mathrm{photons}}{\mathrm{keV\,cm^{2}\,s}}$at\,1\,keV) \dotfill & $ { 644.7 }_{ -50.2 }^{ +55.1 } $ &  $ {  1.66 }_{ -0.17 }^{ + 0.21 } $ &  $ {  3.14 }_{ -0.39 }^{ + 0.48 } $ \\
                  & $\Gamma_{\rm pl}$ \dotfill & $ {  2.35 }_{ -0.05 }^{ + 0.05 } $ &  $ {  2.46 }_{ -0.10 }^{ + 0.12 } $ &  $ {  1.94 }_{ -0.10 }^{ + 0.14 } $ \\
\noalign{\medskip}
        & Statistic/d.o.f.\,$^{\rm \bf c}$ \dotfill & $ 1113 $/$ 1429 $ ($ 0.779 $) &  $ 554 $/$ 477 $ ($ 1.162 $) &  $ 225 $/$ 193 $ ($ 1.168 $) \\
        & Data bins \dotfill & $ 1434 $ &  $ 482 $ &  $ 198 $ \\
        & Net counts\,$^{\rm \bf d}$ \dotfill & $ 55936 $ &  $ 27815 $ &  $ 9592 $ \\
\noalign{\medskip}
                   & $F_{\rm X}$($10^{-14} \mathrm{erg}\,\mathrm{s}^{-1}\,\mathrm{cm}^{-2}$)\,$^{\rm \bf e}$ \dotfill & $    5425.6^{+463.7}_{-422.5}$   &  $    46.40^{+5.87}_{-4.75}$ &  $    155.03^{+23.70}_{-19.26}$ \\
\noalign{\smallskip}         
                   & $L_{\rm X}$($10^{40} \mathrm{erg}\,\mathrm{s}^{-1}$)\,$^{\rm \bf e}$ \dotfill & $    210.3^{+18.0}_{-16.4}$    &  $    1.80^{+0.23}_{-0.18}$ &  $    6.01^{+0.92}_{-0.75}$ \\
\noalign{\smallskip}
\hline
\noalign{\smallskip}
\end{tabular}
\end{center}
\begin{tablenotes}
\item[1]\textbf{Notes.} The components \textsf{apec}, \textsf{tbnew2}, and \textsf{zpowerlw} are in the rest frame of NGC\,4845 ($z=0.003589$). 
$^{\rm \bf a}$~The total Galactic hydrogen mean column density from the UKSSDC archive.
$^{\rm \bf b}$~The \textsf{apec} normalization factor (cm$^{-5}$) is defined as $10^{-14}/(4\pi [D_{\rm A}(1+z)]^2) \int n_{\rm e} n_{\rm H} dV$, 
where $D_{\rm A}$(cm) is the angular diameter distance, $n_{\rm e}$(cm$^{-3}$) the electron density, $n_{\rm H}$ the hydrogen density, and $dV$ the volume element.
$^{\rm \bf c}$ The spectral model was fitted to the \textit{XMM-Newton} spectrum using chi-square ($\chi^2$) and \textsf{mpfit}, whereas C-stat and \textsf{subplex} were used to model the \textit{NICER} data because of low source counts and intense backgrounds. 
$^{\rm \bf d}$ Source net counts were estimated over 0.3--10\,keV after background subtraction and non-X-ray flare removal.
$^{\rm \bf e}$~The unabsorbed X-ray fluxes and luminosities ($D=18$\,Mpc) correspond to the energy band of 0.4--10 keV. 
\end{tablenotes}
\end{table*}

The parameter values for the best-matched models are listed in Table~\ref{ngc4845:model:param}. The best-fit solutions and the corresponding confidence limits at 90\% were derived using the \textsc{isis} standard function for confidence intervals (\textsf{conf\_loop}). For the \textit{XMM-Newton} data, we obtained a spectral index of $\Gamma=2.35\pm 0.05$, which is in agreement with the previous results: $2.36\pm 0.14$ (\textit{Swift}), $2.19\pm 0.03$ \citep[XMM;][]{Nikolajuk2013}, and $2.32 \pm 0.04$ \citep{Perlman2022}. However, modeling the \textit{XMM-Newton} spectrum yielded a column density of $N_{\rm H,pl}=  (11.22 \pm 0.25) \times 10^{22}$\,cm$^{-2}$, which is slightly higher than $(7.21 \pm 0.08) \times 10^{22}$\,cm$^{-2}$ derived by \citet{Nikolajuk2013}. Nevertheless, \citet{Nikolajuk2013} adopted a black body instead of an \textsf{apec} component. Our choice of \textsf{apec}, which is similar to the spectral model employed by \citet{Perlman2022}, is based on the second PCA component found in \S\,\ref{ngc4845:pca}.  However, our \textsf{apec} fitting provided a soft collisional plasma component with a plasma temperature of $kT={ 0.78 }_{ -0.19 }^{ +0.20 }$\,keV, which is higher than the blackbody temperature of $0.33 \pm 0.04$\,keV derived by \citet{Nikolajuk2013} and a collisional plasma temperature of $kT={ 0.24 }_{ -0.05 }^{ +0.15 }$\,keV found by \citet{Perlman2022}. The discrepancies on the column density and collisional plasma temperature between our values and previous values could be related to the different spectral models, background handling, non-X-ray flare removal, and updated calibration data. 

Table~\ref{ngc4845:model:param} also presents the parameter values of the spectral model fitted to the time-averaged \textit{NICER} spectra of all the datasets and those with X-ray flares. It can be seen that the spectral indices of $\Gamma=2.46^{+0.12}_{-0.10}$ and $1.94^{+0.14}_{-0.10}$ were obtained for all combined datasets and those containing X-ray flares, respectively. Therefore, the X-ray source appears to be slightly harder when X-ray flares occur in 2023. The absorbing columns in both sets of \textit{NICER} monitoring observations are much lower than that derived with \textit{XMM-Newton} during the TDE in January 2011, while the column density during X-ray flares captured by \textit{NICER} is also less than that associated with the time-averaged spectrum of all \textit{NICER} data. This means that there was no strong line-of-sight absorbing material in 2023--2024. For the \textsf{apec} component, we obtained plasma temperatures of $kT\approx { 0.17}$ (all datasets) and $0.11$\,keV (X-ray flares), which are much lower than $kT\approx{ 0.78 }$\,keV we obtained for the \textit{XMM-Newton} observation of the TDE in 2011.
Assuming that the plasma temperature derived by \textsf{apec} is the result of a shock-ionized plasma of a colliding wind or jet, the wind velocity can be estimated using the formula: $kT=(3/16)\mu m_{\rm H}v^2$, where $v$ is the wind velocity, $\mu=0.615$ is the mean molecular weight, and $m_{\rm H}$ is the hydrogen mass \citep[see, e.g.,][]{Guedel2009}. For $kT= 0.78^{+0.20}_{-0.19}$\,keV during the TDE, we obtain a wind velocity of $800\pm 100$\,km\,s$^{-1}$. The \textsf{apec} plasma temperatures obtained from the \textit{NICER} observations listed in Table~\ref{ngc4845:model:param} correspond to wind velocities of $380\pm 10$\,km\,s$^{-1}$ (all datasets) and $300\pm 20$\,km\,s$^{-1}$ (X-ray flares). This indicates that there could have been a strong shock wind or jet flow during the TDE in 2011, as previously demonstrated using radio observations \citep{Irwin2015,Lei2016}.

\begin{figure}
\begin{center}
\includegraphics[width=0.443\textwidth, trim = 0 0 0 0, clip, angle=0]{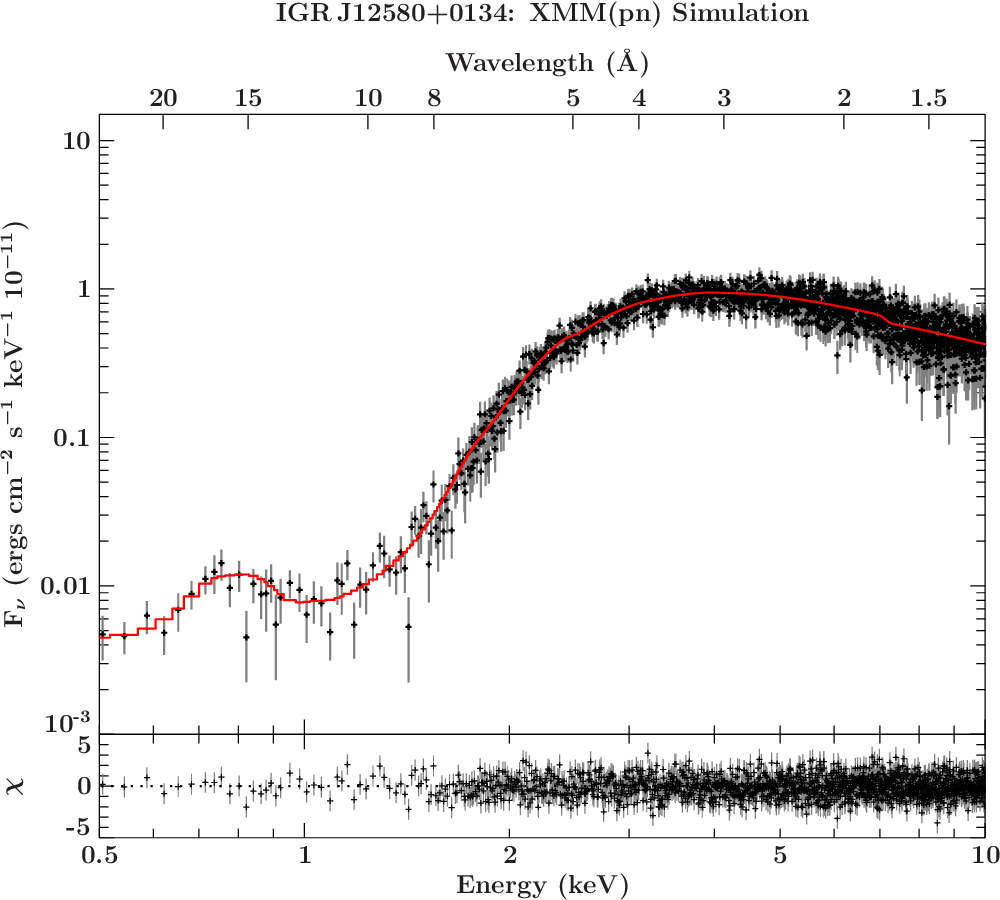}
\end{center}
\caption{Synthetic \textit{XMM-Newton}/EPIC-pn spectrum of X-ray outbursts simulated using the \textit{XMM-Newton} model parameters listed in Table~\ref{ngc4845:model:param}.
\label{ngc4845:fig:sim}
}
\end{figure}

\begin{figure*}
\begin{center}
\includegraphics[width=0.95\textwidth, trim = 0 0 0 0, clip, angle=0]{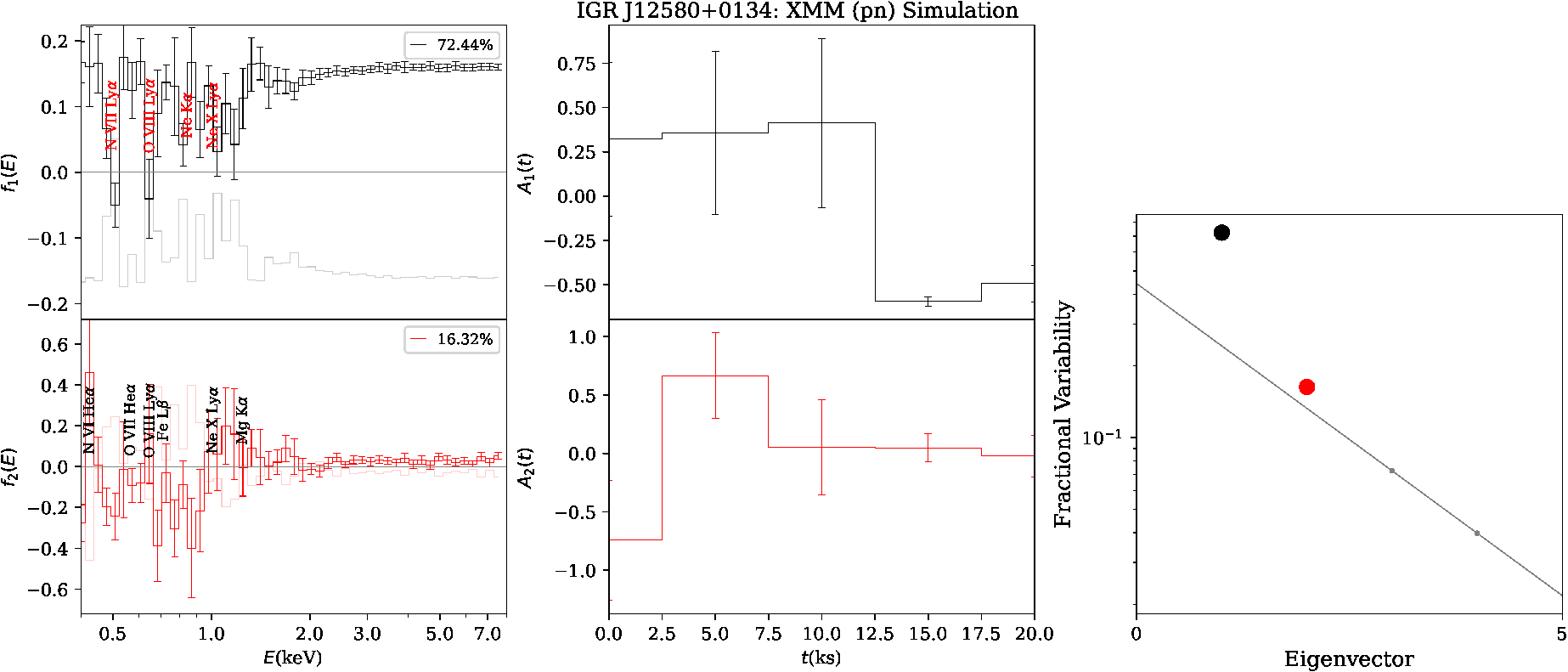}%
\end{center}
\caption{The normalized PCA spectra $f_{i}(E)$ (left), the corresponding light curves $A_{i}(t)$ (middle), and the associated LEV diagram (right) of the synthetic \textit{XMM-Newton}/EPIC-pn spectra simulated at an interval of 5\,ks using the \textit{XMM-Newton} model parameters listed in Table~\ref{ngc4845:model:param} with variable values described in the text.
\label{ngc4845:fig:pca:sim}
}
\end{figure*}

From the model fitted to the \textit{XMM-Newton} observation, we determined an unabsorbed X-ray flux of $F_{\rm X}=5.43^{+0.46}_{-0.42} \times 10^{-11}$ erg\,s$^{-1}$\,cm$^{-2}$ over the energy band of 0.4--10\,keV. Similarly, \citet{Nikolajuk2013} derived an averaged X-ray flux of $6.09 \times 10^{-11}$ erg\,s$^{-1}$\,cm$^{-2}$ over 2--10 keV. In addition, \citet{Perlman2022} estimated a flux of $2.87 \times 10^{-11}$ erg\,s$^{-1}$\,cm$^{-2}$ in the 0.5--7\,keV band. The unabsorbed X-ray flux derived for the \textit{XMM-Newton} spectrum is approximately 120 times higher than that estimated for the time-averaged spectrum of all \textit{NICER} observations. The mean X-ray flux of events with X-ray flares recorded by \textit{NICER} is also elevated by a factor of 3 compared to the average flux of all \textit{NICER} datasets. The X-ray luminosities are also listed in Table~\ref{ngc4845:model:param}, which were estimated by assuming  a redshift distance of $18$\,Mpc from \citet{Yu2022}.

               
\subsection{Simulated X-ray Outbursts}
\label{ngc4845:sim}

To evaluate the feasibility of the PCA-suggested spectral components, we generated simulated variable X-ray spectra for \textit{XMM-Newton}/EPIC-pn utilizing the parameter values of the model fitted to the \textit{XMM-Newton} observation of IGR\,J12580+0134. We performed simulations with the \textsc{isis} function \textsf{fakeit}, incorporating the redistribution and response data from the RMF and ARF of the \textit{XMM-Newton} observation. We used the model \textsf{constant}\,$\times$\,\textsf{tbnew1}\,$\times$\,(\textsf{apec}\,$+$\,\textsf{tbnew2}\,$\times$\,\textsf{zpowerlw}) in which we adjusted the factor in the \textsf{constant} component to reproduce the net count provided in Table~\ref{ngc4845:model:param}. Our simulation approach is similar to \citet{Danehkar2024,Danehkar2024a} in which the X-ray variability was simulated using PCA light curves extracted from observations. Previously, \citet{Koljonen2013} and \citet{Parker2014a} simulated X-ray variations using the \textsc{xspec} command \textsf{fakeit}, albeit with random changes within the confidence limits of the model parameters. We created sets of simulated time-segmented spectra with 5 ks exposure intervals and stored them in PHA files in the FITS file format using Remeis \textsc{isis} functions (ISISscripts). These files were analyzed using the same \textsc{pca} program employed in our analysis in \S\,\ref{ngc4845:pca}.  
 
To simulate the X-ray outbursts, we introduced continuum changes to the simulated time-segmented spectra by multiplying the normalization factors of \textsf{zpowerlw} and \textsf{apec} by $K_{1} (A_{1}-\min(A_{1}))$ and $K_{2} (A_{2}-\min(A_{2}))$, respectively, where $A_{1}$ and $A_{2}$ are the light curves of the first and second PCA components derived from the \textit{XMM-Newton} observation shown in Fig.\,\ref{ngc4845:fig:pca:1}, and $K_{1}$ and $K_{2}$ are arbitrary constants for adjusting the contribution fractions. To better adjust the shapes of our synthetic PCA spectra in the soft exceed ($<2$\,keV), we also added fluctuations to the line-of-sight absorbing material and collisional plasma temperature by multiplying the \textsf{tbnew2} column and \textsf{apec} temperature by $(1 + K_{3} A_{1})$ and $(1 + K_{4} A_{2})$, respectively, where $K_{3}$ and $K_{4}$ are arbitrary constants for adjusting the fluctuation intensities. The well-matched synthetic PCA spectra were reproduced with $K_{1}=2$, $K_{2}=1$, $K_{3}=0.6$, $K_{4}=0.1$. 
The time-averaged spectrum of the simulated X-ray outbursts is shown in Fig.\,\ref{ngc4845:fig:sim}, which is roughly similar to the \textit{XMM-Newton} spectrum in Fig.\,\ref{ngc4845:fig:spec}. Similarly, the simulated spectrum was binned to have at least 15 counts per bin.

The two PCA components derived from our synthetic spectra with the simulated X-ray outbursts are shown in Fig.\,\ref{ngc4845:fig:pca:sim}. These components exhibit spectral patterns similar to those of the normalized PCA spectra shown in Fig.\,\ref{ngc4845:fig:pca:1}. Our analysis indicates that changes in the \textsf{apec} normalization result in a sequence of peaks within the second PCA spectrum, which is analogous to $f_2(E)$ obtained from the \textit{XMM-Newton} observation. However, some discrepancies are seen in the soft excess. It is important to recognize that the characteristics observed in the PCA components of the actual observations may be attributed to an inhomogeneous absorbing column and multi-temperature collisional plasma, whose physical properties varied miscellaneously over the course of the observations. Nevertheless, our simulations conducted using simple assumptions still support the notion that the second component with a relatively time-lagged peak is likely related to collisionally ionized plasma, which may be formed by the interaction of a wind or jet with the surrounding medium, or alternatively, by internal dissipation within the base of a jet \citep[see, e.g.,][]{Lei2016}. 

\vfill\break

\section{Discussions}
\label{ngc4845:discussion}

\subsection{Explanation for the TDE and X-ray Flares}

Our simulations of X-ray outbursts indicate that the brightening event in 2011 was due to a sudden increase in the normalization of the power-law source continuum linked to accretion of the SMBH, as previously argued by \citet{Nikolajuk2013}. Although there is evidence of highly absorbing material during the outbursts, this event was primarily formed by an instant growth in the hard X-ray emission, likely from a corona forming above the accretion flow close to the SMBH, rather than changes in the absorbing material. Similarly, the faint X-ray flares detected by the \textit{NICER} monitoring campaign could be mainly associated with changes in corona emission owing to occasional weak accretion.  

Our spectral analysis shows that the hard continuum of the X-ray source during the TDE can be described by a heavily absorbed power law ($\Gamma\approx2.35$), which agrees with the results of \citet{Nikolajuk2013} and \citet{Perlman2022}. The hard X-ray emission reached its maximum level over a few weeks in January 2011 and then decreased over a year, following a development consistent with a TDE, which was estimated to be from the disruption of approximately 10\% of an object with a mass of 14-30 Jupiter masses \citep{Nikolajuk2013}. However, our regular \textit{NICER} monitoring from March 2023 to February 2024 did not detect any TDE or any decreasing pattern associated with a further TDE, but rather some faint X-ray flares likely from extremely weak accretion.   

\subsection{Soft Thermal Emission from Jet Structures?}

Our simulated X-ray outbursts also support the possibility that the emission features in the soft excess could be from collisional plasma formed by the interaction of a colliding jet or wind rather than thermal emission from a disk. Similarly, \citet{Perlman2022} found that a collisional plasma component modeled by \textsc{apec} provided a better fit to the \textit{XMM-Newton} spectrum, although \citet{Nikolajuk2013} modeled it using a blackbody component. However, our selection of the \textsc{apec} component for modeling soft emissions is based on the second spectral component, which is disclosed by PCA (see Fig.\,\ref{ngc4845:fig:pca:1}).
The normalized light curve of this component also apparently demarcates a lagged peak relative to the peak of the light curve of the first component. While the first component is likely linked to an increase in corona emission as discussed previously, the lagged time series suggests that the second component may not be directly related to corona formation over accretion of the SMBH. Rather, it could originate from the interaction of a jet or wind with the surrounding medium near the SMBH, or from internal dissipation within the base of a jet \citep[e.g.,][]{Lei2016}. 
However, we note that PCA may not naturally preserve time ordering as it decomposes variability to statistically uncorrelated components, which could lead to spurious lags.

Using the Karl G. Jansky Very Large Array (VLA) radio observations at 1.6 and 6\,GHz, \citet{Irwin2015} suggested that the peak X-ray emission in the TDE IGR\,J12580+0134 could be explained by inverse Compton upscattering of the photons produced by relativistic electrons near the center of the galaxy, further establishing the link between self-absorbed synchrotron emission of adiabatic expanding radio outflows and hard X-ray emission of the TDE in galactic nuclei. Moreover, \citet{Lei2016} modeled VLA radio data using a jet model, implying an ``off-beam''\footnote{An ``off-beam'' jet refers to a jet that is not directly pointed toward an observer, while an ``on-beam'' jet is directly aimed at the observer and appears significantly brighter than an off-beam one.} relativistic jet interacting with the circumnuclear medium (CNM), and found that the radio emission can be well described by synchrotron emission from the external shock in the Newtonian regime. Such relativistic jets, which are powered by accretion onto SMBHs, have been found in other TDEs \citep[see review by][]{DeColle2020}, although IGR\,J12580+0134 is the first TDE with an off-beam relativistic jet. Detailed studies of VLA observations also revealed that the time decay of the radio fluxes in IGR\,J12580+0134 was consistent with predictions from a jet--CNM interaction model \citep{Irwin2015,Lei2016,Perlman2017}. Further VLA observations along with reanalysis of previous radio and X-ray observations pointed to a late-time radio flare with a peak on 19 December 2011, which appeared 342 days after the X-ray outburst on 22 January 2011, in addition to a second radio peak in May 2016 \citep{Perlman2022}. However, no X-ray observations were made several months before the second radio peak in 2016; therefore, it is not possible to confirm the presence of a second TDE. The delayed radio flare in IGR J12580+0134 identified by \citet{Perlman2022} was the second such event observed in a TDE, joining ASASSN-15oi detected by \citet{Horesh2021}. 

Because detailed analyses of the VLA radio observations of IGR\,J12580+0134 are consistent with an off-beam relativistic jet, it is possible that the soft thermal emission disentangled by PCA from the X-ray variations is related to the jet structure. Previously, \citet{Lei2016} argued that radio emission corresponds to an external shock in the Newtonian regime, whereas bright X-ray flares of the TDE could be associated with (1) the corona, (2) internal dissipation within the jet, and (3) external shock. They favored the first scenario in which the intense X-ray TDE described by a power law apparently originates from a corona, which is also supported by our PCA study and spectral modeling. However, soft thermal emission does not seem to be directly linked to coronal emission; rather, it originates in the region very close to the corona, which could be in the ambient medium within the proximity of accretion flows to the SMBH or internal dissipation within the base of a jet near the innermost accretion disk around the SMBH.

\subsection{Faint X-ray Flares: Evidence for AGN Activity?}

The \textit{NICER} monitoring observations of IGR\,J12580+0134 revealed the presence of faint X-ray flares in 2023 (see Fig.\,\ref{ngc4845:fig:lc}), which supports the notion that there is more likely a low-luminosity AGN in NGC\,4845. The presence of an AGN in this nearby spiral galaxy has been suggested based on the spectral fluctuations observed in radio- and X-ray observations \citep{Irwin2015,Auchettl2017}. 
Using CHANG-ES, \citet{Irwin2015} identified a variable compact radio core, consistent with the presence of an AGN in NGC\,4845. Moreover, \citet{Auchettl2017} challenged the categorization of IGR\,J12580+0134 as a TDE, citing that the host galaxy NGC\,4845 is indeed a Seyfert 2 or low-ionization nuclear emission-line region (LINER) AGN \citep[as assigned by][]{Ho1995,Veron-Cetty2006} based on its pre-flare optical emission lines, mid-infrared \textit{WISE} colors and hardness ratio evolution, and argued that this could be explained by a changing-look AGN. Moreover, \citet{Perlman2022} acknowledged the presence of AGN activity in this object, but contended that the extremely luminous X-ray event recorded in January 2011, designated as IGR\,J12580+0134, is most likely a TDE for two reasons: (1) the X-ray luminosity and follow-up radio flux in 2011 were significantly above what is typical for an AGN, (2) the hard X-ray flux evolution over 17.3--80\,keV had a time decay with a $t^{-5/3}$ law, as noticed by \citet{Nikolajuk2013}, being fully consistent with a TDE \citep[see review by][]{Saxton2020}, as observed in other TDEs \citep[e.g.,][]{Lodato2011,Kawamuro2016,Onori2019}.  
Some recent analytic studies have also suggested that the time decay of a TDE declines as $t^{-9/4}$ if the star is partially disrupted \citep{Coughlin2019a,Miles2020}. While no TDE was captured by \textit{NICER} monitoring, the faint X-ray flares recorded by \textit{NICER} once again suggest the existence of a weakly accreting SMBH and the possibility of a low-luminosity AGN at the center of the galaxy NGC\,4845.

\section{Summary}
\label{ngc4845:summary}

We have analyzed the \textit{XMM-Newton} and recent \textit{NICER} monitoring observations of the TDE IGR\,J12580+0134 located at the center of the host galaxy NGC\,4845 in the Virgo cluster. The TDE occurred in November 2010 and was subsequently observed by \textit{XMM-Newton} in January 2011. To identify a future TDE based on the two peaks of radio outbursts in 2012 and 2016 reported by \citet{Perlman2022}, we initiated \textit{NICER} monitoring for approximately one year from March 2023 to February 2024, leading to the detection of some faint X-ray flares. To investigate the spectral state transitions during the TDE and X-ray flares, we have conducted a hardness analysis of the light curves. We have used PCA to disclose the spectral components appearing during X-ray outbursts and flares, which were then employed to decide the spectral models for describing the time-averaged spectra, as well as the simulations of X-ray outbursts. We present a summary of the key findings as follows:

(i) The X-ray source observed by \textit{XMM-Newton} is exceptionally luminous with random fluctuations during the TDE in January 2011, but extremely dim in June 2021. In addition, the \textit{NICER} light curves showed X-ray brightening flares from March to June 2023, followed by a period of reduced luminosity in 2024. However, the hardness ratio ${\rm HR}_{H}$ has high levels during the TDE in 2011, in addition to the slightly elevated hardness ratio ${\rm HR}_{M}$. In contrast, both hardness ratios remained roughly stable with random fluctuations throughout the \textit{NICER} monitoring campaign during 2023--2024.

(ii) Our analysis of the \textit{XMM-Newton} observations using PCA showed two distinct spectral components. The first PCA spectrum corresponds to a heavily absorbing power-law spectrum in the TDE, as its corresponding light curve increases when the TDE occurs. The second PCA component is likely associated with thermal emission features originating from collisionally ionized plasma, likely caused by a colliding wind or jet, whose light curve demonstrates an apparently lagged peak relative to that of the first component. PCA of the \textit{NICER} campaign similarly showed the same spectral components, albeit with much less absorption of the continuum emission.   

(iii) Our spectral analysis of the time-averaged spectrum collected by \textit{XMM-Newton} during the TDE implies an extremely bright power-law source with an unabsorbed 0.4--10 keV flux of $\approx 5.4 \times 10^{-11}$ erg\,s$^{-1}$\,cm$^{-2}$  and a photon index of $\Gamma \approx 2.35$, similar to previous results \citep{Nikolajuk2013,Perlman2022}, which is heavily absorbed by the line-of-sight material with a column density of $\approx 11.2 \times 10^{22}$\,cm$^{-2}$. However, the source is less obscured over the course of the \textit{NICER} observations, but by a factor of 120 fainter than that during the TDE. The unabsorbed flux of the time-averaged \textit{NICER} spectrum of the source when X-ray flares occurred was approximately three times higher than that of all the combined \textit{NICER} observations. In addition, the spectral model fitted to the \textit{XMM-Newton} observation of the TDE  provides us with a collisional plasma temperature of $\approx 0.78$\,keV, which is higher than $\approx  0.17$\,keV obtained from the \textit{NICER} time-averaged spectrum. If the collisional plasma is created in a shock-ionized region, such a plasma temperature of $0.78$\,keV is the result of a wind moving with $800$\,km\,s$^{-1}$.

(iv) The synthetic spectra simulated using the spectral model fitted to the \textit{XMM-Newton} observation of the TDE indicated that increases in the normalization of the power-law component during the TDE led to the first principal component determined from PCA. In addition, the column density of the material blocking the power-law continuum was found to be higher during the TDE. Moreover, our simulations of the \textit{XMM-Newton} spectra also showed that changes in the normalization of an \textsc{apec} component could result in a second PCA component associated with collisionally ionized plasma in the soft excess. 

X-ray observations of IGR\,J12580+0134 suggest a potential link between TDEs and ongoing low-level AGN activity. NGC\,4845 was previously classified as a Seyfert 2/LINER galaxy \citep{Veron-Cetty2006}, indicating the presence of an AGN. A bright TDE occurred in late 2010 \citep{Walter2011}, with a late radio flare in late 2011, followed by another radio event in the middle 2016 \citep{Irwin2015,Perlman2022}. Although it is plausible that AGN activity mimics a TDE \citep{Auchettl2017}, other criteria such as late radio flares and X-ray light-curve decay ($t^{-5/3}$) favor the presence of a TDE in 2010 \citep{Perlman2022}. The \textit{NICER} monitoring observations revealed some faint X-ray flares in 2023 that were much weaker than the original TDE, which could be attributed to variations in the accretion rate, so a recurrent partial TDE is unlikely to have occurred during the \textit{NICER} observations. Our findings highlight the need for long-term monitoring of host galaxies of TDEs to understand the possible connection between TDEs and AGN activity.

In conclusion, although the \textit{NICER} monitoring campaign of IGR\,J12580+0134 did not detect an additional TDE, these observations revealed the presence of faint X-ray flares, likely produced by extremely weak accretion onto the SMBH at the center of the spiral galaxy NGC\,4845, suggesting that this galaxy may host a low-luminosity AGN.



\begin{acknowledgments}
This research was supported by the NASA/Goddard Space Flight Center through the NICER mission (proposal ID
6093) under grant 80NSSC23K1098, and has made use of data and/or software provided by the High Energy Astrophysics Science Archive Research Center (HEASARC), a service of the Astrophysics Science Division at NASA/GSFC; the \textsc{sas} software for XMM-Newton; a collection of \textsc{isis} functions (ISISscripts) provided by ECAP/Remeis Observatory and MIT; and \textsf{Astropy}, a community-developed core Python package for astronomy. Based on observations obtained with XMM-Newton, an ESA science mission with instruments and contributions directly funded by ESA Member States and NASA.
The author thanks the anonymous referee whose suggestions improved the manuscript.
\end{acknowledgments}

%

\vspace{1mm}


\software{\textsc{xspec} \citep{Arnaud1996}, \textsc{isis} \citep{Houck2000}, \textsf{NumPy} \citep{Harris2020}, \textsf{matplotlib} \citep{Hunter2007}, \textsf{Astropy} \citep{AstropyCollaboration2013}.}

\facilities{XMM (pn), NICER.}




{ \small 
\begin{center}
\textbf{ORCID iDs}
\end{center}
\vspace{-5pt}

\noindent A.~Danehkar \orcidauthor{0000-0003-4552-5997} \url{https://orcid.org/0000-0003-4552-5997}

}





\end{document}